\documentclass[12pt]{article}\usepackage[hyperfootnotes=false]{hyperref}
\usepackage{epsfig}

\usepackage{floatrow}
\usepackage{caption}

\usepackage{amsmath}
\usepackage{amssymb}
\usepackage{graphicx}
\newlength{\abstractwidth}
\renewcommand{\thefootnote}{\fnsymbol{footnote}}
\renewcommand{\thanks}[1]{\footnote{#1}} 
\newcommand{\starttext}{
\setcounter{footnote}{0}
\renewcommand{\thefootnote}{\arabic{footnote}}}
\renewcommand{\theequation}{\thesection.\arabic{equation}}
\newcommand{\be}{\begin{equation}}
\newcommand{\bea}{\begin{eqnarray}}
\newcommand{\eea}{\end{eqnarray}}
\newcommand{\beq}{\begin{equation}}
\newcommand{\ee}{\end{equation}}
\newcommand{\eeq}{\end{equation}}

\renewcommand{\a}{\alpha}

\def\ba{\begin{eqnarray}}
\def\ea{\end{eqnarray}}

\def\q{{1\over 4}}
\def\12{{1 \over 2}}
\def\eq{&=&}

\def\ra{\rangle}

\def\simleq{\; \raise0.3ex\hbox{$<$\kern-0.75em
\raise-1.1ex\hbox{$\sim$}}\; }
\def\simgeq{\; \raise0.3ex\hbox{$>$\kern-0.75em
\raise-1.1ex\hbox{$\sim$}}\; }

\def\s{{\cal{S}}}
\def\O2{\Omega_2}

\def\bi{\begin{itemize}}
\def\ei{\end{itemize}}

\def\sc{\setcounter{equation}{0}}

\def\W{$\Omega$}
\def\W'{$\Omega$}

\def\V{\Omega}
\def\V'{\Omega}

\def\a{{\cal{A}}}
\def\O{${\cal{O}}$}

\def\c{{\cal{C}}}
\def\p{{\cal{P}}}
\def\q{{\cal{Q}}}

\def\s{{\cal{S}}}

\def\bn{\bigskip \noindent}

\def\lds{l}
\def\suk{SU(2^K)}
 \def\cp{CP(2^K-1)}
 \def\cft{conformal field theory}
  \def\cfts{conformal field theories}
  
    \def\cg{$\c$-geometry}
     \def\cg2{$\c_2$-geometry}

\makeatletter
\g@addto@macro\normalsize{%
  \setlength\abovedisplayskip{10pt}
  \setlength\belowdisplayskip{20pt}
  \setlength\abovedisplayshortskip{10pt}
  \setlength\belowdisplayshortskip{20pt}
}
\makeatother

\usepackage{color}

\begin{document}
\renewcommand{\theequation}{\thesection.\arabic{equation}}
\begin{titlepage}
\rightline{}
\bigskip
\bigskip\bigskip\bigskip\bigskip
\bigskip
\centerline{\Large \bf {Switchbacks and The Bridge to Nowhere  }}

\bigskip
\begin{center}
\bf  Leonard Susskind and Ying Zhao \rm

\bigskip

Stanford Institute for Theoretical Physics and Department of Physics, \\
Stanford University,
Stanford, CA 94305-4060, USA \\
\bigskip

\end{center}

\begin{abstract}

This paper is in three parts: Part 1 explains the relevance of Einstein-Rosen bridges for one-sided black holes. Like their two-sided counterparts, one-sided black holes are connected to ERBs whose growth  tracks the increasing complexity of the quantum state. Quantitative solutions for one-sided ERBs are presented in the appendix.

Part 2 calls attention to the work of Nielsen and collaborators on the geometry of quantum  complexity. This geometric formulation of complexity provides a valuable tool for studying the evolution   of complexity for  systems such as black holes.

Part 3 applies the Nielsen approach to geometrize  two related black hole quantum phenomena: the rapid mixing of information through fast-scrambling; and the time dependence of the complexity of  precursors, in particular the switchback effect.

\medskip
\noindent
\end{abstract}

\begin{figure}[h!]
\begin{center}
\includegraphics[scale=.5]{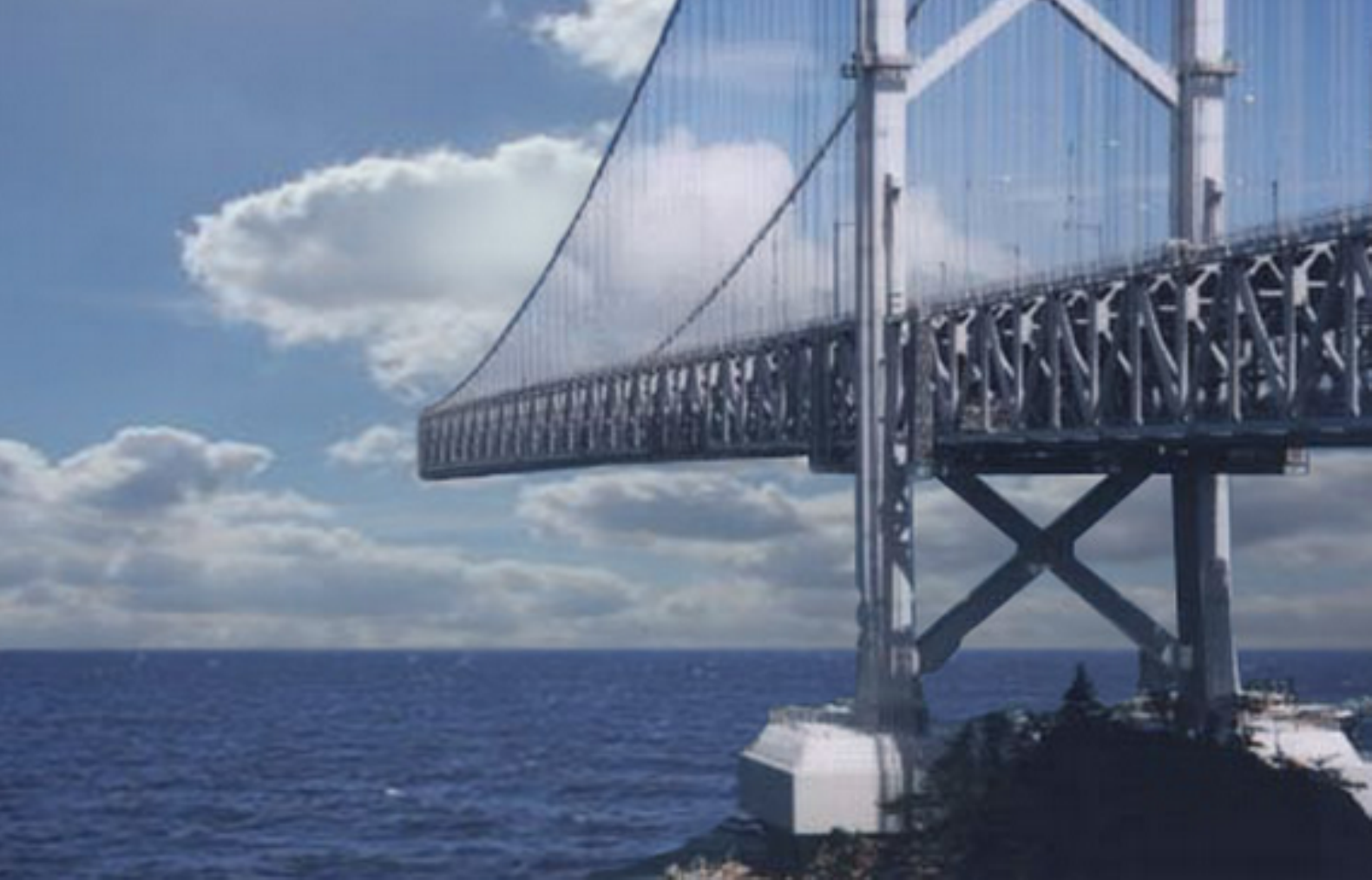}
\caption*{ }
\label{}
\end{center}
\end{figure}

\end{titlepage}

\starttext \baselineskip=17.63pt \setcounter{footnote}{0}
\tableofcontents

\sc
\section{Introduction}\label{Introduction}

In reference \cite{Stanford:2014jda}  evidence was given for the duality  between two time-dependent black hole phenomena:

\bi

\item the growth of the volume of an Einstein-Rosen bridge (ERB) for a two-sided ADS black hole.

\item the growth of computational complexity of the two-sided state of the dual gauge theories.

\ei

Although it has not been widely discussed, the concept of an ERB applies equally well to one-sided black holes as to the two-sided case. Furthermore the duality between ERB-volume and computational complexity may also be formulated for the one-sided case.

The plan of this paper is to first  generalize the concept of an ERB to the case of a one-sided black hole in a pure state. At first sight such a generalization seems contradictory;  ERBs are ordinarily considered to be   the geometric connection formed between entangled systems according to the ER=EPR conjecture. Because a black hole in a pure state is not entangled with anything else it would seem that there is no such thing as a one-sided ERB. Nevertheless we will see that there is a very definite sense in which the one-sided black hole is attached to a  ``bridge-to-nowhere."  Moreover, as in the two-sided case, the bridge-to-nowhere grows with time, tracking the growth of  complexity.

The second direction that we explore has to do with the definition of complexity. Modeling black hole evolution by random quantum circuits \cite{Hayden:2007cs} has been a useful guide for some purposes. But black holes are not random circuits; they evolve by Hamiltonian evolution. The definition of complexity in terms of quantum gates may be ideal for computer science, but not for black holes.

In an attempt to be quantitative about gate complexity, Nielsen and collaborators \cite{Nielsen}\cite{Dowling} constructed a continuum approximation to gate complexity which involves a new kind of ``complexity geometry." Thus far the Nielsen geometrization of complexity has not proved useful for quantum-computer science, but it is just what we need for black holes. We will illustrate the value of the geometric approach by applying it to an interesting aspect of black hole evolution: the switchback effect \cite{Stanford:2014jda} associated with precursors and shockwave geometries \cite{Shenker:2013yza}. The switchback effect is also closely related to fast-scrambling \cite{Hayden:2007cs}\cite{Sekino:2008he}.

The methods we describe apply to both  one and two sided cases, but for definiteness we assume a single copy of a gauge theory and a one-sided black hole.

\sc
\section{The Bridge to Nowhere}\label{S bridge to nowhere}

According to the ER=EPR hypothesis of \cite{Maldacena:2013xja} an ERB is the wormhole that connects a black hole to whatever it is entangled with. The most studied case is the two-sided ADS black hole (really a pair of black holes), dual to a pair of entangled \cfts \cite{Maldacena:2001kr}\cite{Hartman:2013qma}. When viewed from the outside the black holes on either side are stationary, but the ERB is strongly  time-dependent. Its volume increases with time, and appears to track the growing complexity of the \cft \ state.

As we will explain, a one-sided black hole although unentangled, is attached to a growing  `bridge-to-nowhere. To minimize new terminology we'll use ERB1 and ERB2 to indicate one and two-sided ERBs.

\begin{figure}[h!]
\begin{center}
\includegraphics[scale=.3]{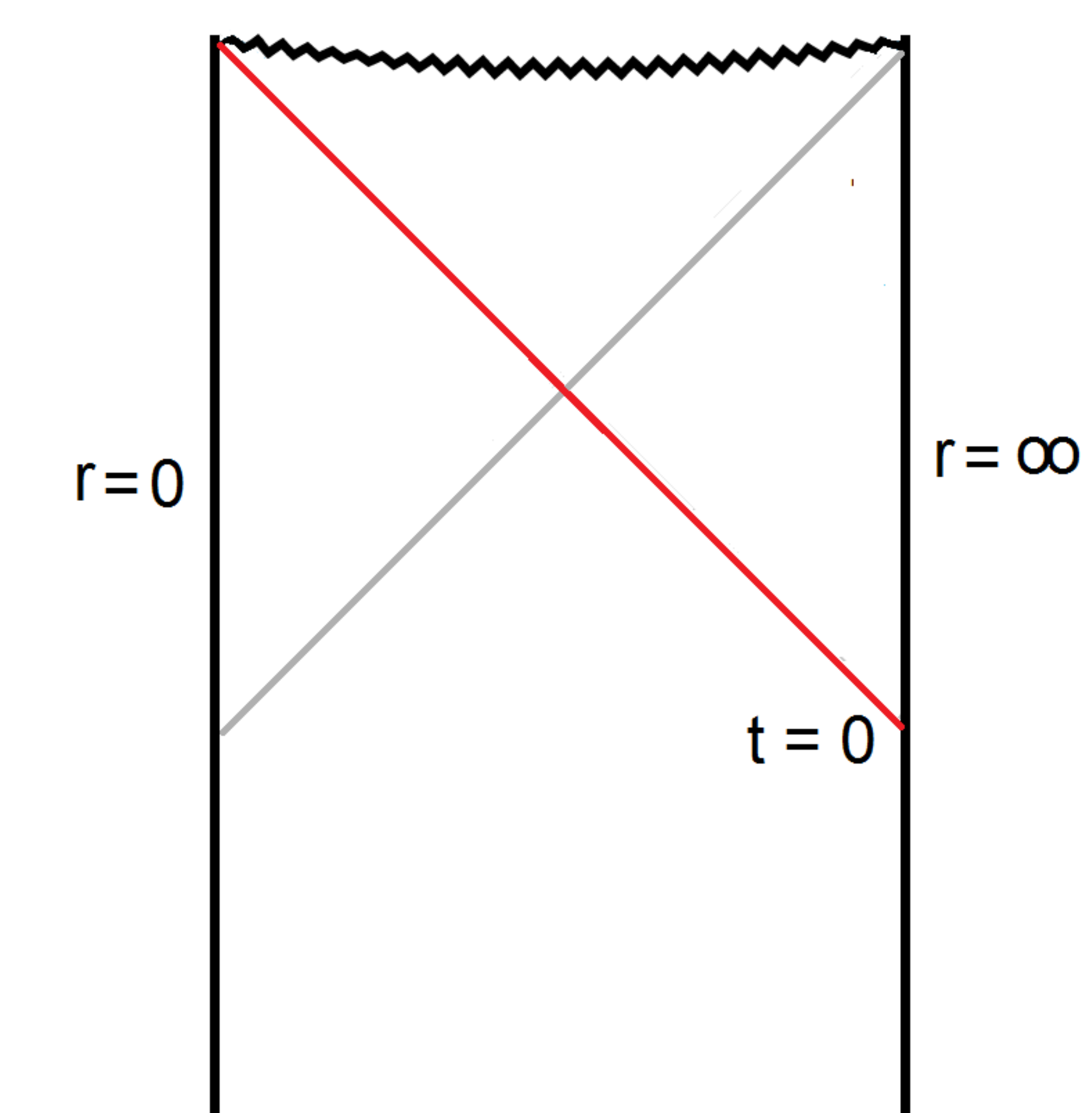}
\caption{Penrose diagram for one-sided ADS black hole. The red line is an infalling shell. }
\label{1}
\end{center}
\end{figure}

To define an ERB1 we begin with a \cft  \ on a spatial sphere cross time---$S(D-1) \times R.$ Initially the system is in the ground state dual to the ADS vacuum. At $t=0$ a  source is activated  which injects enough energy to create a stable black hole. The most tractable situation is when the matter comes in as a spherically symmetric light-like shell.
 The Penrose diagram in figure  \ref{1} shows the process.

\begin{figure}[h!]
\begin{center}
\includegraphics[scale=.3]{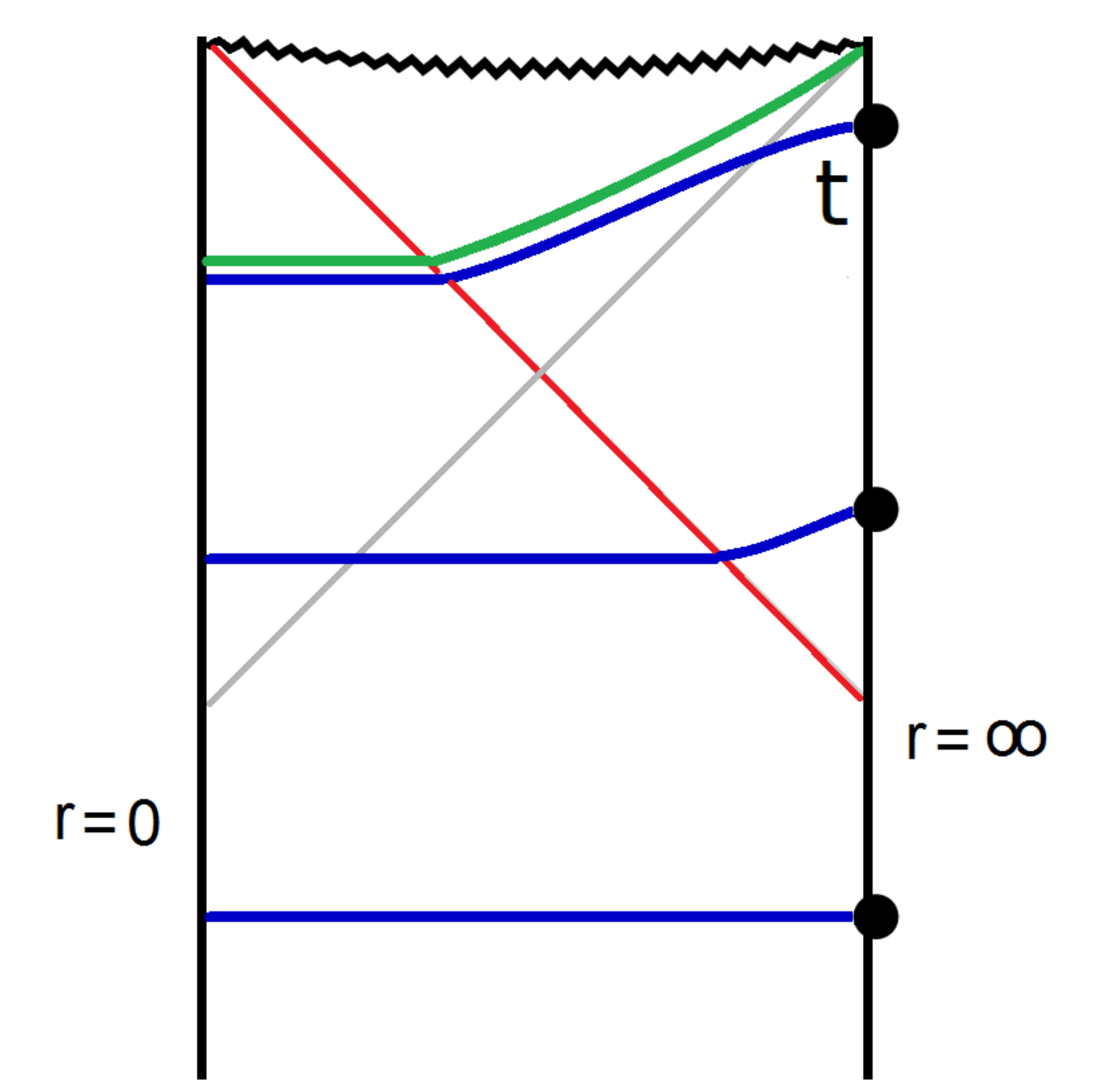}
\caption{The blue curves show maximal volume spatial surfaces anchored at the boundary. The green curve is the asymptotic late-time limit of the ERB. }
\label{2}
\end{center}
\end{figure}

The ERB1 is defined by analogy with the two sided construction of \cite{Stanford:2014jda}. Pick a time $t$ on the boundary. This defines a boundary $(D-2)$-sphere. Then consider the family of all $(D-1)$-dimensional volumes (space-like hypersurfaces which end on that sphere). The surface of maximal volume defines the ERB1 at time $t.$ In figure \ref{2} the evolution of the ERB1 is shown as a series of blue spacelike curves which intersect the boundary at various times.

The equations governing the maximal surface are the same as for the two-sided black hole \cite{Stanford:2014jda} and the results are similar. The details are in the Appendix.

The spatial volume of these maximal surfaces is infinite for the obvious reason that the metric diverges at the ADS boundary. That divergence can easily be regulated, either by cutting off the integrals at some  radial regulator coordinate or by subtracting the vacuum ADS value. Another way to accomplish the same thing is to only count the volume behind the horizon.

\begin{figure}[h!]
\begin{center}
\includegraphics[scale=.3]{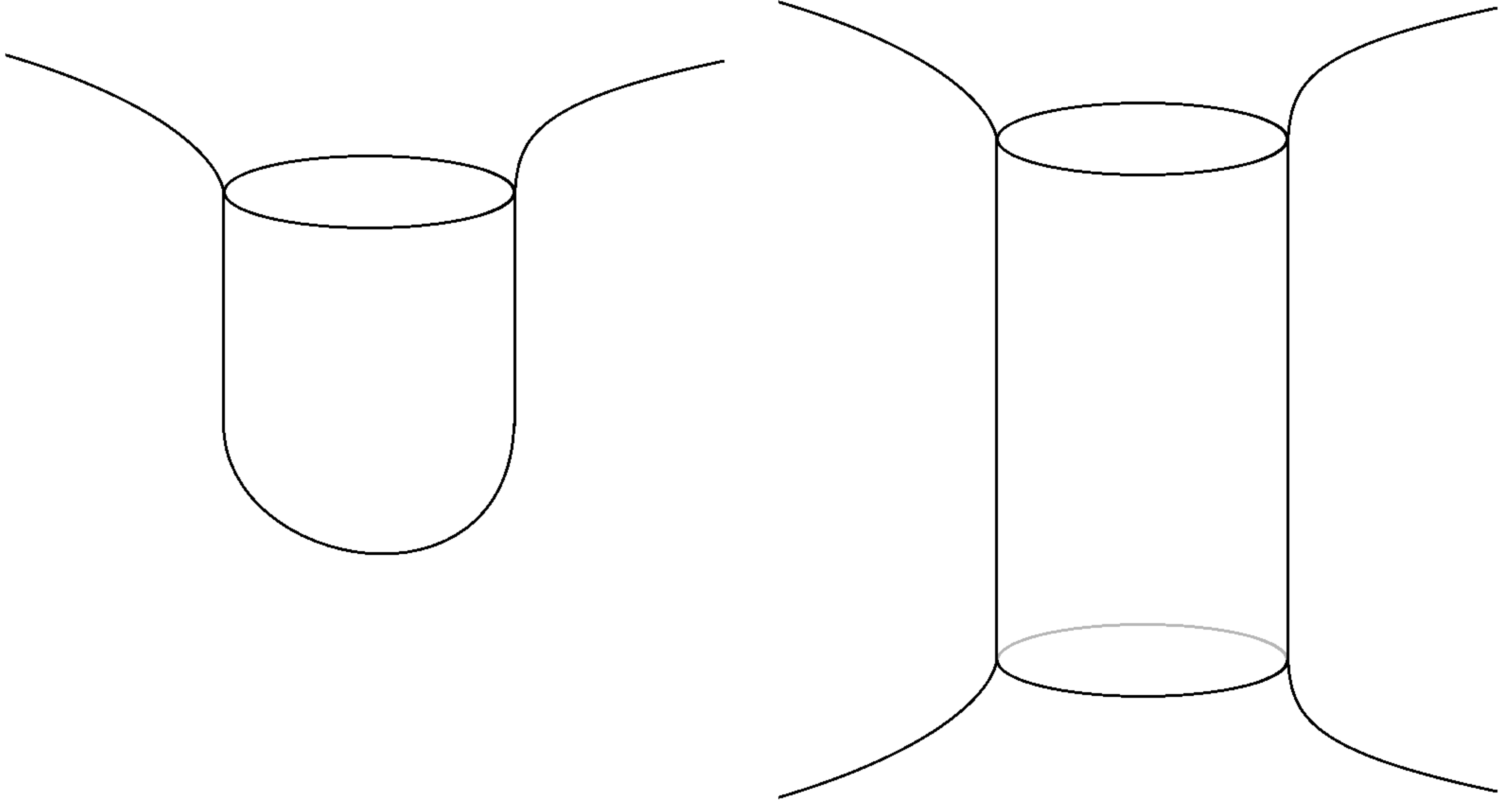}
\caption{ Embedding diagrams of one and two-sided ERBs.}
\label{3}
\end{center}
\end{figure}

As in the ERB2 case, as $t\to \infty $ the ERB1 asymptotically approaches a fixed  hypersurface of infinite regulated volume, show as the green curve in figure \ref{2}. For finite $t$ it grows by a process of stretching which takes place just inside the horizon.
Figure \ref{3} shows embedding diagrams for both the two and one-sided ERB's at some particular time.

A worthwhile point to note is that if there are two unentangled black holes then the embedding diagram is just two copies of the left side of figure \ref{3} with no internal connection. The right side of the figure is an example of the slogan ``Entanglement provides the hooks that hold space together." In this case entanglement stitches the two one-sided ERBs together to form a two-sided ERB.

For $t<0$ the ERB1 is merely a constant time slice through ADS. In general the portion of the ERB1 inside the shell is part of such a constant time slice, but after it passes through the shell it becomes modified. Once the black hole has formed the volume of the ERB1 grows linearly with time,

\be
V=A_{erb} \ t
\ee
 with $A_{erb}$ being the cross-sectional area. Within a numerical factor discussed in \cite{Stanford:2014jda} $A$ is the area of the horizon.

The computational complexity of the state also grows linearly with a rate equal to the entropy $S$ times the temperature $T$ of the black hole. Calling the complexity $\c,$

 \bea
 \c(t) &\approx& STt \cr \cr
 \eq \frac{A_{erb}}{4G}T t \cr \cr
 \eq V \frac{T}{G}
 \eea

The example that was discussed at length in \cite{Stanford:2014jda} was the minimal stable black hole of Schwarzschild radius equal to the ADS radius of curvature $l.$

\be
R=l
\label{R}
\ee
 and temperature,

\be
 T=1/l.
 \label{T}
 \ee
In that case

\be
\c \to \frac{V}{G \lds}
\ee
From here on the term black hole will refer to  one satisfying \ref{R} and \ref{T}\footnote {Larger black hole in ADS will be discussed in a upcoming paper with Dan Roberts and Douglas Stanford  \cite{RSS}.}

\sc
\section{The Geometry of Complexity} \label{S geometry complexity}

\subsection{Operator Complexity}

Classically the growth of the ERB (either one or two-sided) takes place forever. Quantum mechanically  the global classical geometry must break down by the quantum recurrence time, and probably by the classical recurrence time\footnote{The quantum recurrence time is doubly exponential in the entropy while the classical recurrence time is given by a single exponential.}. These are extremely long times; much longer than any ordinary thermalization time or evaporation time. The question is what quantity in the dual gauge theory continues to grow for such long periods? The answer is quantum computational complexity. In this section we will review some  geometric ideas about complexity which are due to Michael Nielsen and collaborators \cite{Nielsen}\cite{Dowling}, and then apply them to fast scrambling, and to the switchback effect of \cite{Stanford:2014jda}.

A quantum circuit  (QC) is a  device composed of qubits and gates (called $g$) whose purpose is to implement specific unitary transformations on an initial state of the qubits. Let the number of qubits be $K.$ The gates are one or two-qubit special-unitary transformations\footnote{By special-unitary we mean with determinant $1.$}. The gates can either occur in series---in each time step a single gate acts---or in parallel as explained by Hayden and Preskill \cite{Hayden:2007cs}. In the Hayden-Preskill parallel case the qubits are paired at each step, and $K/2$ gates act simultaneously on the pairs. For our purposes in this section the distinction is not important and it is easier to describe the series case.

\begin{figure}[h!]
\begin{center}
\includegraphics[scale=.4]{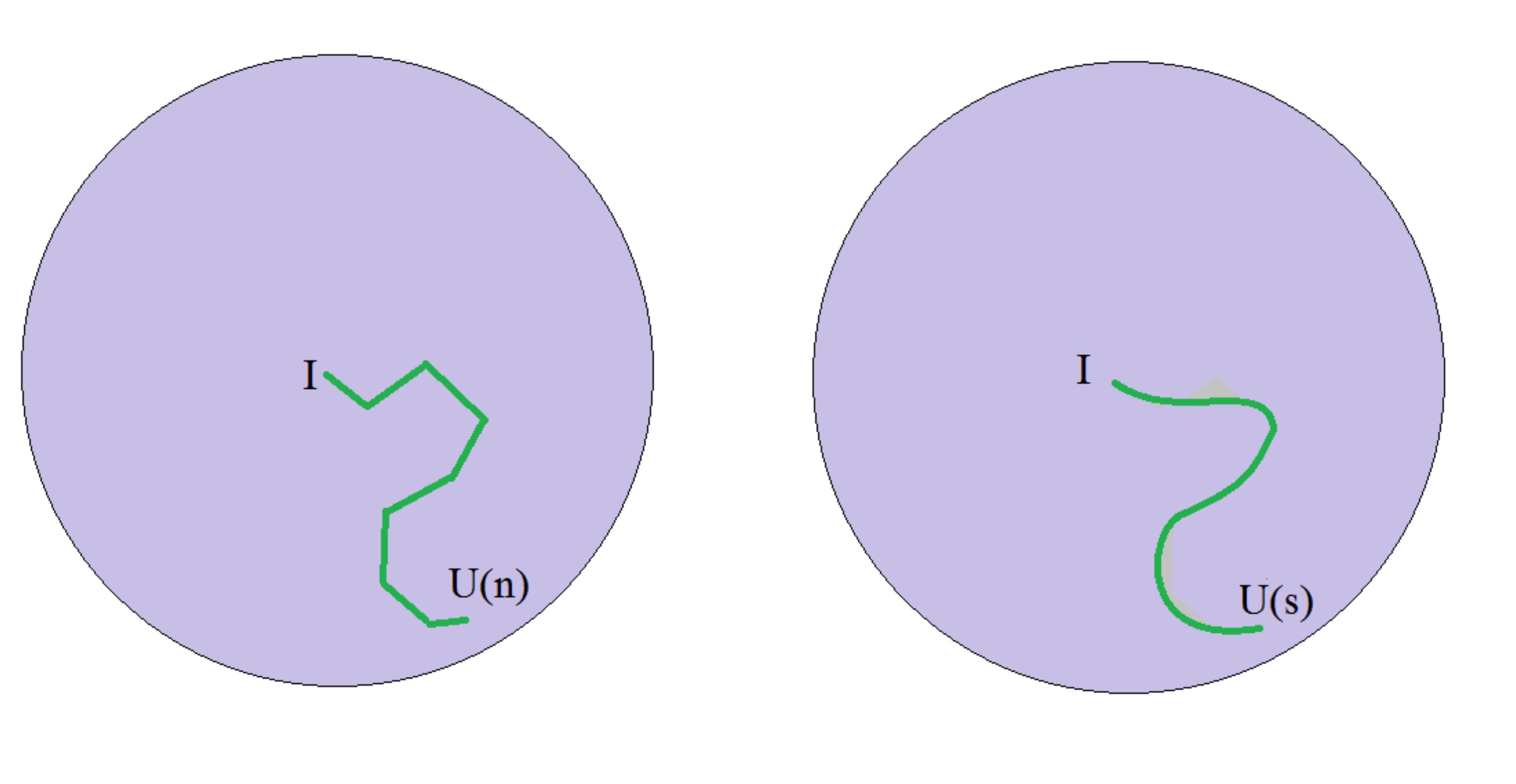}
\caption{ The shaded area represents the group manifold $\suk.$} The left side shows a discrete path induced by a series of gates. The right side shows a curve induced by a Hamiltonian evolution.
\label{4}
\end{center}
\end{figure}

Given an initial state $|\Psi(0)\ra$ after $n$ time-steps the state has been transformed to $|\Psi(n)\ra,$

\bea
|\Psi(n)\ra \eq g_n ....g_3 g_2 g_1  |\Psi(0)\ra \cr \cr
\eq u(n) |\Psi(0)\ra.
\label{gpath}
\eea
Here $u(n)$ is an element of $SU(2^K).$ Let us think of \ref{gpath} as defining a path in the space $SU(2^K).$  This is schematically shown in the left side of figure \ref{4}.
The path begins at $u(0) = I$ and ends at $u(n).$  The rule for such paths is that every link corresponds to a gate and therefore displaces the endpoint by a one or two qubit operator.

With these concepts in hand we can define the complexity of a unitary operator $u$ as the smallest number of gates of any circuit that can yield $u$ as an outcome. That is to say, it is the number of links of the shortest allowable path connecting $I$ and $u.$

 In the past, random quantum circuits have been used to model black hole evolution \cite{Hayden:2007cs}\cite{Stanford:2014jda}
 but our real interest is in continuous Hamiltonian evolution. Part of the reason for this paper is to draw attention to an
innovation of Nielsen  and collaborators \cite{Nielsen}\cite{Dowling} who introduced a continuum description of complexity.  Their purpose was to construct an approximation to a quantum circuit that used Hamiltonian evolution and Riemannian geometry. However, the methods of  \cite{Nielsen}\cite{Dowling}  seems  well suited to the study of  Hamiltonian systems of the kind that may represent black hole evolution.

 What the authors of \cite{Dowling} do is to introduce a Riemannian geometry on the space $SU(2^K).$ The discrete time steps are replaced by infinitesimal transformations and the discrete path through $\suk$ by a continuous path. This is shown on the right side of figure \ref{4}.

The path may be parameterized by a generator $h$ that acts as  a possibly path-dependent Hamiltonian\footnote{The notation $h$ is used to represent any generator composed out of the $Z_A$ basis of operators. The notation $H$ is reserved for the actual Hamiltonian governing the black hole evolution. Similarly the notation $u(s)$ is used for a generic trajectory on $\suk$ and $U$ is used for the actual evolution induced by $H.$}.

\bea
\frac{d u}{ds} \eq -ih(s) u  \cr \cr
u(s) \eq P e^{-i\int_0^s h(s^{\prime})ds^{\prime}}
\eea
where $P$ indicates path-ordered product. Here $s$ is a path parameter that we'll think of as time, and $h(s)$ is a time dependent Hamiltonian. The construction yields a path through $SU(2^K)$ where at every point the tangent to the path is determined by $h(s).$  It is natural in this setting to define the complexity of $u$ to be the length of the shortest geodesic connecting $I$ and $u.$

 The only question is how to choose the metric on $SU(2^K).$ There is a conventional metric  on unitary groups that is invariant under conjugation $u \to \ vuv^{\dag},$ but it does not have the right properties for defining complexity. What one wants is a right-invariant metric that incorporates a rule that all gates should be  one and two qubit operators. The idea is to choose the geometry so that a high price (in length) is paid for all infinitesimal motions that are not generated by one and two qubit Hamiltonians. More precisely a large price should be paid for Hamiltonians which are not sums of one and two qubit terms.

Let us begin with the algebra of all Hermitian operators constructed from $K$ qubits. Each qubit has four  operators consisting of three traceless Pauli operators and the unit operator. We label the four operators with an index $i$ running over four values.  Let $X^i(n)$ be the four operators for the $n^{th}$ qubit ($i= 1, .., 4$). The algebra is spanned by a basis   consisting of products of $K$ factors,  one $X$ for each qubit. There are $4^K$ basis operators of this type. If we want to consider only traceless operators then there the basis contains $4^K-1$ operators that we will indicate by $Z_A.$ The index $A$ runs over all  $4^K-1$ orthogonal directions in the algebra.

Let us consider a point $u_0$ in $SU(2^K).$ Consider the displacements

\be
\delta_A  = -i Z_A u
\label{displace}
\ee
defined by left multiplication by $Z.$ The conventional $\suk$ metric is equivalent to the following inner product structure.

\bea
\delta_A \cdot \delta_B \eq \frac{1}{2^K} \mathrm{Tr} Z_A Z_B \cr \cr
\eq \delta_{AB}
\eea
It is invariant with respect to  right multiplication by unitary matrices and defines a homogeneous  geometry. It also has a discrete form of isotropy. Any two basis generators  $Z_A$ and $Z_B$ can be related by conjugation by an $SU(2^K)$ operation. However, in general normalized superpositions of the basis generators are not equivalent. An easy way to see that is to note that any $Z_A$ satisfies

\bea
\mathrm{Tr} Z_A  \eq 0 \cr \cr
Z_A^2 \eq 1
\eea
but superpositions of the generators do not satisfy the second equation for any normalization.

However this metric is  not quite what we want. We would like  a more  anisotropic  geometry in which displacements are very costly
for  directions $Z_A$ corresponding to three or more qubit operators. To that end \cite{Dowling} introduces projection operators $\p$ and $\q$  and a cost-parameter $q>>1.$ The projector $\p$ projects onto the subspace of one and two qubit $Z$-operators. Those directions are called ``easy" directions and no large cost is attached to them. The projector $\q$ projects onto the orthogonal ``hard" subspace of three and more qubit operators. The $\c$-metric is then given by,

\be
\delta_A \cdot \delta_B = (\p +q\q)_{AB}
\label{comp metric}
\ee
If $A$ and $B$ are both easy then the inner product is $\delta_{AB}.$ If they are both hard the inner product is $q \delta_{AB}.$ Otherwise the inner product is zero. For $q\gg 1$ the metric defined by \ref{displace} and \ref{comp metric} was called the ``standard" metric in \cite{Dowling}. We will call it the $\c$-metric to avoid confusing it with other more conventional metrics. When $q$ is large the $\c$-metric introduces a large price in metrical distance for curves whose tangents project onto hard directions. The geometry of $\suk$ defined by the $\c$-metric will be referred to as $\c$-geometry.

 $\c$-geometry is homogeneous but when $q$ is large it is extremely anisotropic. Geodesics strongly avoid hard directions.  Reference \cite{Dowling} shows that if  $q>4^K,$ then the complexity is insensitive to its precise value\footnote{The limit $q\to \infty$ was discussed in \cite{Nielsen}.}.

We may now define the complexity of an element $u$ of $SU(2^K).$ Using the $\c$-metric with large $q$ the complexity is defined by
\it the  distance between $I$ and $u$ along the shortest geodesic connecting them. \rm

There is a very important case which the authors of \cite{Dowling} study in detail. If $H$ is a fixed easy Hamiltonian\footnote{Here we have in mind the actual Hamiltonian governing the evolution of the the black hole. }, i.e., a Hamiltonian composed as a sum of one and two qubit terms, independent of $s,$ then the curve

\be
u(s) = e^{-ihs}
\label{curve}
\ee
is a geodesic. In other words time-independent easy Hamiltonians generate geodesics. Moreover for $s$ smaller than some critical value $s_c$ the curve in \ref{curve} is the shortest geodesic connecting $I$ and $u(s).$ The length of such a geodesic is proportional to $s$ which shows that for some length of time the complexity grows linearly with time. This was assumed in \cite{Stanford:2014jda}  but in \cite{Dowling} it is proved.

Beyond the critical time $s_c$ equation \ref{curve} continues to define a geodesic but it is not the shortest one. The transition to a shorter geodesic occurs at or before the first  conjugate point at a so-called \it cut point\rm\footnote{There may be more than one geodesic connecting two points in a Riemannian space. As one proceeds along the geodesic generated by \ref{curve}, at some point a second geodesic may replace it as the shortest one. Such a point is called a \it cut point \rm.   At a conjugate point a continuous family of geodesics connect two points. If a conjugate point exists then a cut point must occur at or before the conjugate point. }. One can see that a cut point must happen for sufficiently large $s.$ The quantum recurrence theorem implies that that $e^{-iHs} \approx I$ for $s\sim e^{2^K}.$ Thus for very long times the geodesic \ref{curve} cannot be the shortest.

This double exponential time is probably overkill. One knows that there is a maximum complexity for circuits. It is of order $e^K$
beyond which the complexity cannot increase. Therefore after  a singly exponential  time (classical recurrence time), shorter geodesics must replace \ref{curve}. But there is no general argument that requires that they occur before the classical recurrence time. In previous work we assumed that complexity increases linearly until it reaches is maximal value $e^K.$ It is important to know whether this is true for the kinds of Hamiltonians which can describe black holes.

An important fact noted in \cite{Dowling} is that most sectional curvatures of the $\c$-geometry are negative. This means that neighboring geodesics tend to deviate exponentially. The qualitative reason is not too hard to understand; most classical Hamiltonians are chaotic which means that trajectories have a  tendency to exponentially separate. This is also a property of geodesics in a negatively curved space. We will return to this in the  next section on switchbacks.

\subsection{State Complexity}

Thus far we have discussed the complexity of unitary operators. One can also define complexity for states. To do so we need to first define simplicity. For  a system of qubits that's fairly easy. Intuitively a state is simple if it not entangled. In other words product states are simple. The simplest of all product states is one in which all the qubits are all in the same state. For example the state

\be
|0\ra \equiv |0000,....,0\ra
\ee
is simple.

An arbitrary normalized state $|\Psi\ra$ may be identified with a point on the homogeneous  manifold
$\cp.$

 The complexity of a general state $|\Psi \ra$ can be defined to be the complexity of the least complex unitary that can act on $|0\ra$ to give $|\Psi \ra.$

 Let's see if we can define a metric on $\cp.$ Consider two neighboring points, i.e., normalized vectors in the qubit Hilbert space modulo overall phase. Call them $|f\ra$ and $|g\ra.$ It is assumed that the difference is infinitesimal. There exist unitary transformations which connect the two states. The unitary operators are not at all unique but among them there is a unitary of minimal complexity. This means that it is the shortest distance from the identity using the complexity metric. We take the complexity of that unitary to define the distance between the two points $f, \ g$ on $\cp.$

 It is not hard to prove that the defining properties of a metric are satisfied:

 \bea
 d(x,y) & \geq & 0 \cr \cr
 d(x, y) \eq 0  \ \ \ \ \rm if \ and \ only \ if \ \it  x = y  \cr \cr
  d(x, y) \eq  d(y,x) \cr \cr
 d(x, z) & \leq & d(x, y) + d(y, z) \ \ \ \ \rm (triangle \ inequality ) \it
 \eea

 We can now define the complexity of any state to be the minimum geodesic distance from the simple state to the state in question.

The choice of simple state is not unique.  One could choose any unentangled state instead of
  $$
|0\ra \equiv |0000,....,0\ra
$$
However all such states are close because one can go from one to another with at most $K$ gates. In metric terms they are all within distance $K$ from one another. Distance $K$ sounds large but in the context of the parallel computing model of Hayden and Preskill it corresponds to a single time-step.

The complexity of quantum states can be vastly larger than the complexity of classical states. As a model for a classical system let's consider $K$ c-bits.
Each c-bit is like a coin, it can be in one of two states---heads or tails. We will add the rule that states are identified under the global $Z_2$ that flips all coins.

The state $(hhhhh....hhh) = (ttttt...ttt)$ is a simple state; you can describe it in two words, \it all equal. \rm  Instead of quantum gates we can allow configurations to be updated by simple single-coin flips---heads goes to tails; tails goes to heads. The complexity of a configuration like $(htthtthhhththhhtt...)$ may be defined as the minimum number of flips needed to get to it from the simple state. Here are two  facts:

The maximum complexity of any classical state is $K/2.$ This is obvious; any configuration can be achieved by just flipping the initial configuration to the desired final configuration---one coin at a time.

The second fact is almost as obvious. It says that when $K$ is large almost all states have complexity close to $K/2.$

To put it another way, if one imagines cycling though the states by some rule that flips one coin at a time (the rule for which coin is flipped may depend on the configuration of the other coins) in a reversible manner, then it takes a time of order $K/2$ to reach a maximally complex state. The complexity will hover around $K/2$ for a long time but eventually come back to simplicity in a time $2^K.$

Something like that is true for quantum complexity except the magnitudes are much larger. First of all, within a given accuracy any state can be achieved using $\sim e^K$ gates. Secondly, almost all states have complexity of order $e^K.$ Finally, it typically takes an exponential time to get to most states from the simple state and a doubly exponential time to get back to a simple state. The set of states that you can get to in polynomial time is very special and not at all generic.

Now that we have an understanding of complexity we can state the conjecture relating complexity to the geometry of the black hole interior.

\bn

 \it For a black hole of ADS radius and entropy $K$  the volume of the ERB1 is proportional to the complexity of the quantum state. \rm

 \bn

 More precisely

\be
\c = \frac{V}{G l}
\label{c=V/Gl}
\ee

A more general formula can be based on the following hypothesis. The rate of growth of complexity (long before the recurrence time) should reflect the number of qubits and the energy per qubit. This translates to the rate

\be
\frac{d\c}{dt} = ST
\ee
where $S$ and $T$ are the entropy and temperature of the system. Writing $S = A/4G$ one finds

\be
\c =\frac{ At T}{G} = V \frac{T}{G}
\ee
To specialize this general formula to the case $R=$ we recall that the temperature in that case in $T=1/l$ giving
\ref{c=V/Gl}.

This relation has been confirmed for the TFD state and a wide variety of shockwave deformations of the TFD.

\sc
\section{The Geometry of Precursors and Switchbacks}\label{precursors and switchbacks}

\subsection{Size and Complexity of Precursors}

The section makes heavy use of ideas due to Dan Roberts and Douglas Stanford which will appear shortly \cite{RSS}.

 Precusors are  useful probes that have been used to test the connection between complexity and ERB \cite{Stanford:2014jda}. In this section we will use the geometric complexity ideas of \cite{Nielsen}\cite{Dowling} to analyze them.
Precursors,  and switchbacks are related concepts that apply to products of operators taken out of time order. A precursor is an operator of the form

\be
W(t) = U(-t) W U(t)
\label{precursor}
\ee
where $U(t)$ is the time evolution operator for an arbitrary system. In practice this will mean black hole satisfying \ref{R} and \ref{T}.
The reason they are interesting is that their  complexity evolves in a very characteristic manner. Moreover when a precursor acts on a black hole, the evolution of precursor complexity closely parallels the evolution of the ERB. Our goal in this section is to relate the properties of precursors to properties of the $\c$-geometry. We will then have three distinct ways to evaluate the complexity of precursors: two old and one new. The old ways are: using Einstein field equations to calculate the response of the ERB geometry to the action of a precursor \cite{Shenker:2013yza}; and using random quantum circuits to evolve the precursor. The new way is to study the geodesics of  $\c$-geometry. We already know that the old ways agree with each other \cite{Stanford:2014jda}. With some simple assumptions about the $\c$-metric we will see that the new way  agrees with the old.

Two quantities will occupy our attention; the first is the complexity of the precursor labeled $\c_W(t).$ The other is the  \it size \rm of the precursor \cite{RSS}. Roughly speaking it is the a measure of how strongly the other qubits are causally influenced by the qubit $W$ after time $t.$

The Pauli operators  of the $n$th qubit will be called $X_i(n).$ For most purposes the subscript $i$ can be left out without creating ambiguities. The qubit defining the precursor can be arbitrarily identified as $W=X(1).$  The causal effect of $W(t)$ on another qubit $X_n$ may be defined by the commutator

$$
 [X_n(t), X_m(0)].
$$

The magnitude of the effect is given by the norm of the commutator,

\bea
C(1,n;t) \eq \frac{1}{2^K} \bf Tr \it  [X_1(t), X_m(0)]^2 \cr \cr
\eq  \frac{1}{2^K} \bf Tr \it  [W(t), X_m(0)]^2
\label{Cwn}
\eea

More generally we can define,

\be
C(m,n;t) = \frac{1}{2^K} \bf Tr \it  [X_m(t), X_m(0)]^2
\label{Cmn}
\ee

One way to define the size of the precursor is by,

\be
s_{w}(t) = \sum_{ n} C(1,n;t).
\label{sum}
\ee

We can also define a symmetric version of the size by summing over all pairs of qubits,

\be
\s(t) = \sum_{m\neq n} C(m,n;t).
\label{double sum}
\ee

\subsection{Quantum Circuit Analysis}

We are interested in understanding the complexity and size of precursors by relating them to properties of the $\c$-geometry, but it is helpful to first get an idea of how they work in the Hayden-Preskill  quantum circuit  (QC) model. We begin with the size $s_w$.

In the Hayden-Preskill model, in each time step $K/2$ gates act. The qubits are randomly paired and a gate acts on each pair. In the next step the pairing is randomized again and another $K/2$ gates act. The time steps are taken to be thermal times equal to the inverse temperature. For the black holes we are considering the time steps are the ADS time $l.$ Thus after time $t,$ $t/l$ steps have occurred and $Kt/2l$ gates have acted.

At time $t=0$ all qubit operators for different qubits commute. After one time-step, the first qubit  $W=q_1$ has paired with one other qubit (call it $q_2$), and has interacted with it via a two-qubit gate. Therefore $W$ no longer commutes with
$q_2,$ but it commutes with all others.  Let's say that         $W$ has \it infected \rm $q_2.$  Thus after one step there are two infected qubits.

In the next step the two infected qubits interact with new partners. If we ignore the small probability (of order $1/K$) that $W$ and $q_2$ interact again with each other, then four qubits will become infected, meaning that they will not commute with $W.$ Moreover the non-zero commutators will be of order unity.

Clearly the epidemic of qubits infected by $W$ grows exponentially for a time.
The size of the precursor---the number of infected qubits---will be denoted by $s_w.$ Since an infected qubit has an order one commutator with $W$ it follows that the size is of order

\be
s_w(t) \sim \sum_m C(1,m;t)=  2^{-K} \sum_m  \rm Tr \it [X_1(t),X_m(0 )]^2.
\label{1qubit contribution}
\ee

We can compute how $s_w$ grows in the QC model as follows: Suppose at any given instant the number of infected qubits is $s$ with $s < K.$ In the next time step the average number of newly infected qubits will be,

\be
\Delta s = \frac{K-s}{K-1}s.
\ee
$s$ is given by the cumulative sum of the $\Delta s$ up to time $t.$

A single step of the quantum circuit corresponds to a time of order $\Delta t \sim l.$
Smoothing out the time steps  gives a differential equation for $s(t),$

\be
l\frac{ds}{dt} = \frac{K-s}{K-1}s.
\label{diffeq}
\ee

For $s\ll K$ and $K \gg 1$ we see that $s(t)$ grows exponentially with $t/l.$ But once $s\sim K $ the growth saturates. The solution to \ref{diffeq} is,

\be
s= \frac{Ke^{\frac{t}{l}}}{K-1 + e^{\frac{t}{l}}}.
\label{s from epidemic}
\ee

Notice that the saturation takes place at about the scrambling time. Beyond that $s$ tends quickly to its asymptotic value $s\to K.$

The size $s$ is  one term in the double sum \ref{double sum} that defines $\s(t).$ Therefore we multiply $s$ by $K$ to get,

\be
\s= \frac{K^2e^{\frac{t}{l}}}{K-1 + e^{\frac{t}{l}}}.
\ee

 $\s$ asymptotically tends to the value $K^2.$

 Now let us consider the complexity. After time $t$ the number of gates that has acted  to form $U(t)$ is $p= \frac{K}{2}\frac{t}{l}.$ We may write,

 \be
U(t) = g_p \ g_{p-1}........g_1.
\ee

 The precursor is given by

 \be
W(t) = g^{\dag}_1.......g^{\dag}_{p-1} \ g^{\dag}_p \ W \ g_p \ g_{p-1}........g_1
\label{gate def}
\ee

All together $2p+1$ gates have acted (We may consider $W$ to be a one-qubit gate).
But the total number of gates is an over-estimate of the complexity for the
simple reason that fewer gates would have gotten us to the same result for $W(t).$ It is easy to see that only the gates that were involved in the spread of the epidemic were required, all the others having canceled in \ref{gate def}. This implies that the complexity is given by  the sum\footnote{We thank Dan Roberts and Douglas Stanford for explaining this point.} of the sizes for all time-steps earlier than $t.$ We may write this in differential form,

\be
\c(t) = \frac{ 1}{l} \int_0^t dt \ s(t)
\label{integrating}
\ee

or

\be
s(t) = l\frac{d\c(t)}{dt}
\label{s=dc/dt}
\ee

or in terms of the quantity $\s,$

\be
\s(t)= lK\frac{d\c(t)}{dt}
\ee

For large time

\be
\s(t) \to K^2.
\label{s to K2}
\ee

The complexity
can be obtained by integration using \ref{s from epidemic} and \ref{integrating}. We find,

\be
\c = K \log{\left(1+\frac{1}{K}e^{t/l}\right)} = K \log{\left(1 + e^{\frac{(t-t_{\ast})}{l}}\right)}
\label{C from QC}
\ee

where $t_{\ast}=l\log{K}$ is the scrambling time.

\subsection{Switchback Effect}\label{switchback effect}

at large time \ref{C from QC} behaves like\footnote{There is a tricky factor of $2$ here. This expression is actually twice the complexity of the operator $U(t).$ The origin of the factor of $2$ is the fact that $K/2$ gates occur in a unit time step.}

\be
\c(t) \to K\frac{t}{l} -K\log{K} =  \frac{K}{l}(t-t_{\ast})
\label{swb efft}
\ee

The subtraction of $K\frac{t_{\ast}}{l}$ in \ref{swb efft} is the switchback effect that was
explained in \cite{Stanford:2014jda}. Let us take a moment to recall its importance. One might have thought that the precursor's effect on the volume of an ERB would be proportional to the total number of gates that act in the product $U(t) W U(-t).$ But complexity is not
defined by the number of gates in a construction like equation \ref{gate def}. It is defined by the \it minimum \rm number of gates in any quantum circuit that can produce  $W(t).$ Any gate in $U(t)$ that does not act on an infected qubit might as well not have occurred since it will cancel with a corresponding gate in $U(-t).$ Thus these gates should be deleted in the complexity book keeping.

What is remarkable is that exactly the same cancelation occurs  \cite{Stanford:2014jda} when computing the effect of a Shenker-Stanford shockwave on
 the volume of the ERB. Although the shockwave was created by acting with a precursor, the effect is  a General Relativity effect having nothing obvious to do with quantum complexity.

Later we will see that exactly the same effect occurs in the Nielsen geometric way of thinking about complexity.

Finally we remark that for small $t$ both $s$ and $\c$ grow exponentially,
\be
\c \sim s \sim e^{t/l}  \ \ \ \ \ \ (t<<l \log{K})
\ee

\begin{figure}[h!]
\begin{center}
\includegraphics[scale=.3]{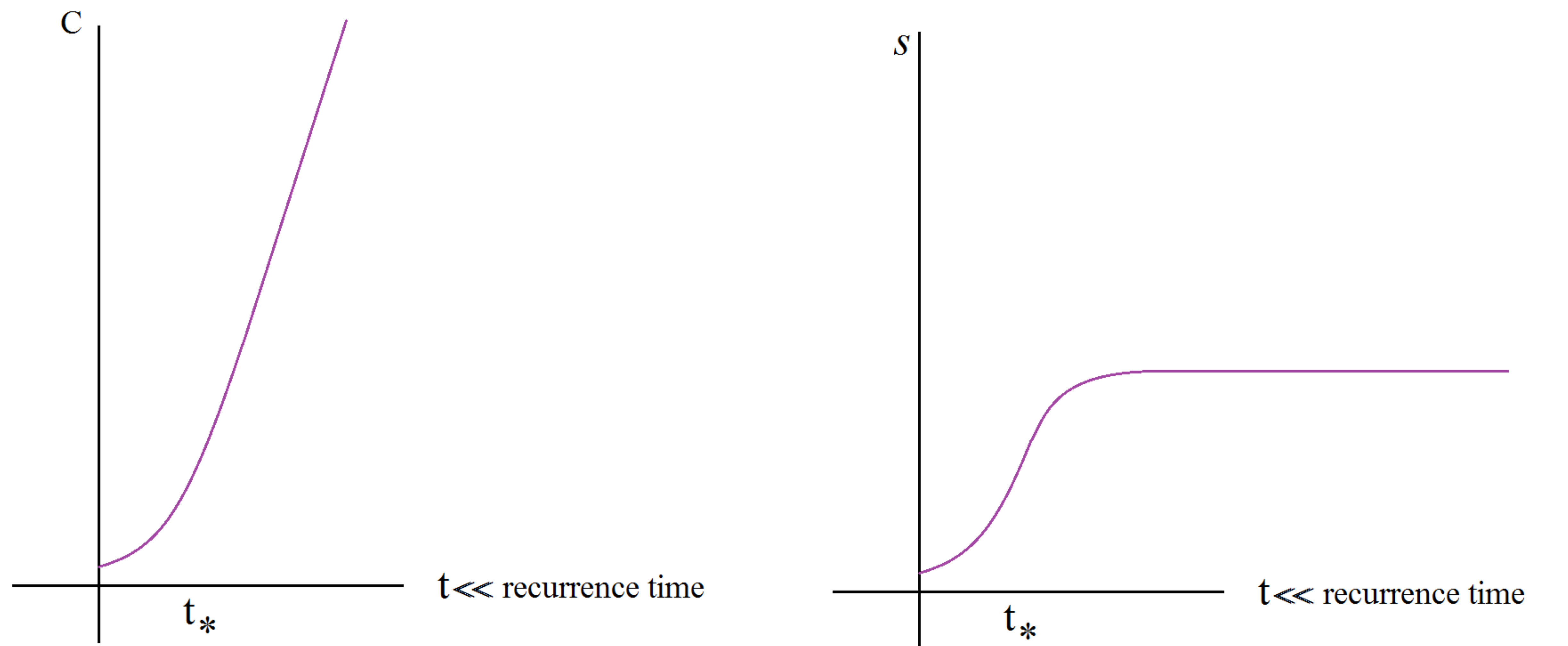}
\caption{Graphs of the complexity and size of a precursor.}
\label{5}
\end{center}
\end{figure}

The qualitative behavior of $\c$ and $s$ are sketched in figure \ref{5}.

\subsection{Complexity-geometry of Precursors}

Our goal is to understand how the properties of complexity and size relate to  $\c$-geometry.

The precursor \ref{precursor} defines a path through the $\c$-geometry that begins with the
 evolution operator $U(s) = e^{-iHs}$  sweeping out a geodesic connecting the unit operator with $U(t).$ We will refer to it as a ``vertical" geodesic\footnote{The terminology vertical and horizontal are just names with no technical significance.}. Next, $W$ acts to add a short segment of unit length in a new ``horizontal" direction.  Finally $U^{\dag}$ acts and continues the curve more or less backward vertically along the direction of $U.$

\begin{figure}[h!]
\begin{center}
\includegraphics[scale=.3]{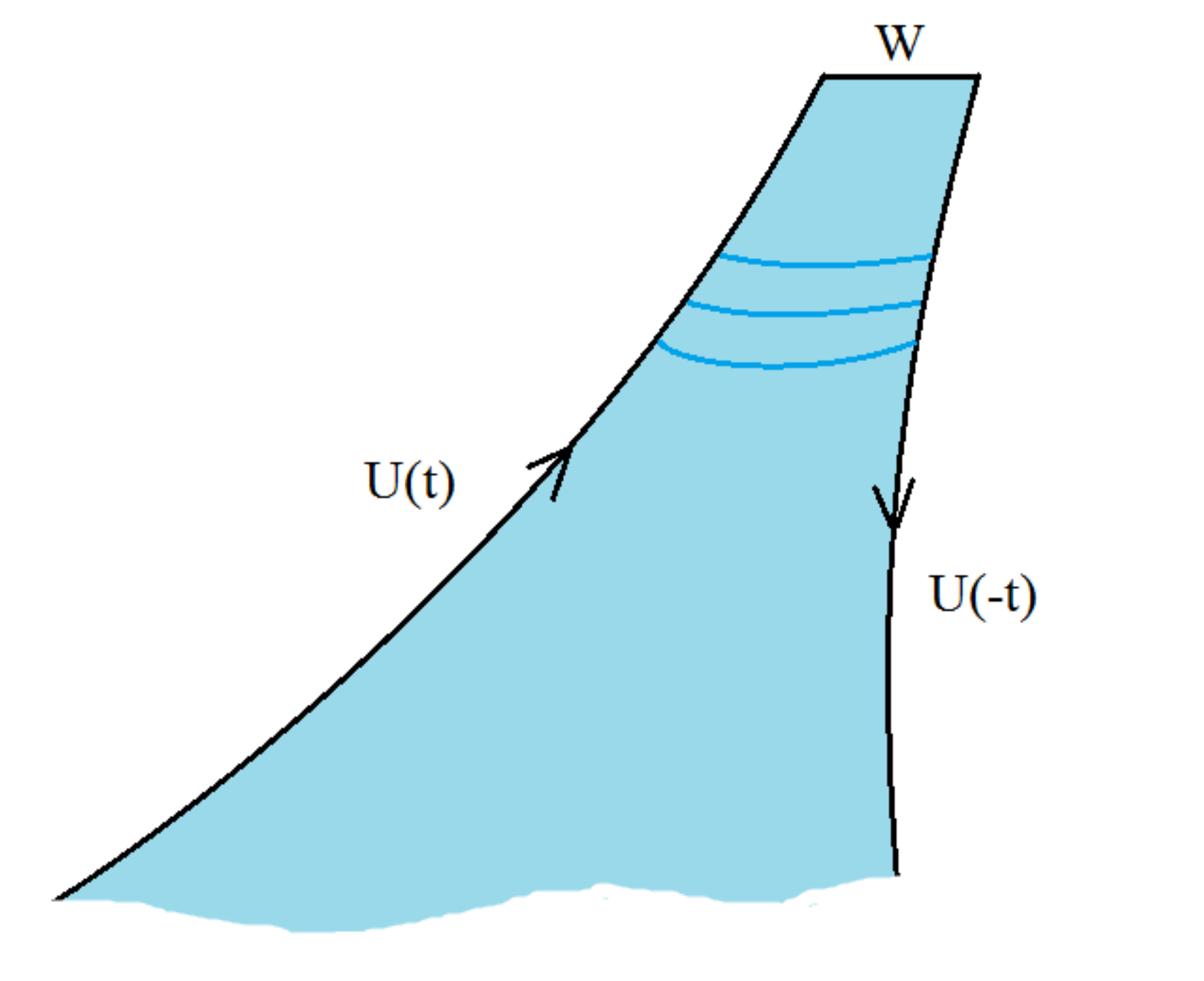}
\caption{The black trajectory shows the ``vertical" geodesics $U(\pm t)$ connected by a short ``horizontal" segment $W.$ The blue curves are horizontal geodesics connecting $U(t)$ with $U(-t).$ As $t$ varies the blue curves sweep out a two dimensional surface shown in light blue. }
\label{6}
\end{center}
\end{figure}

 We will now define a two dimensional sub-manifold of  $\c$-geometry \  associated with the precursor: call it a $\c_2$-geometry.  $U(t)$ and $U(-t)$ define segments of neighboring geodesics connected by a third short segment representing the one-qubit operator $W.$ This is depicted in figure \ref{6}.

Next, `` horizontal" geodesics are constructed connecting corresponding points along $U(\pm t).$ As $t$ is varied from $t=0$ to $t=\infty$ the horizontal geodesics sweep out a surface with boundaries. This surface is the  $\c_2$-geometry for the precursor \ref{precursor}.  Its metric is inherited from the original $\c$-metric defined in section \ref{S geometry complexity}. Note that the length of the horizontal geodesics represent the complexity of the precursor at time $t.$

In principle, given a Hamiltonian and a $W$ it is possible to construct the metric on the two-dimensional geometry, but this would be very hard. Our goal will more modest; namely we will do a bit of guessing about the metric and then attempt to verify the properties of complexity and scrambling that we indicated earlier.

First consider whether the two-dimensional $\c_2$-metric is positively or negatively curved. It it were positively curved the geodesics $U(\pm t) $ in figure \ref{6} would tend to meet after a finite time. This is impossible because the geodesic  generated by a given Hamiltonian, and passing through a given point, is unique. Moreover, if the Hamiltonian leads to chaotic behavior, as we expect for a black hole, then the trajectories should exponentially diverge. This suggests that the curvature of the
$\c_2$-metric is negative.  Our first  guess is that the curvature is negative and constant over the the portion of the $\c_2$-geometry of interest. As for the magnitude of the curvature we could leave it as a parameter to be fixed by comparison with \ref{switchback effect}. This is an exercise that we will leave to the reader. The result is that the  two-dimensional curvature scalar $R_2,$ must be chosen to satisfy,

\be
R_2 = -\frac{4}{K^2}.
\label{2curvature}
\ee

\begin{figure}[h!]
\begin{center}
\includegraphics[scale=.3]{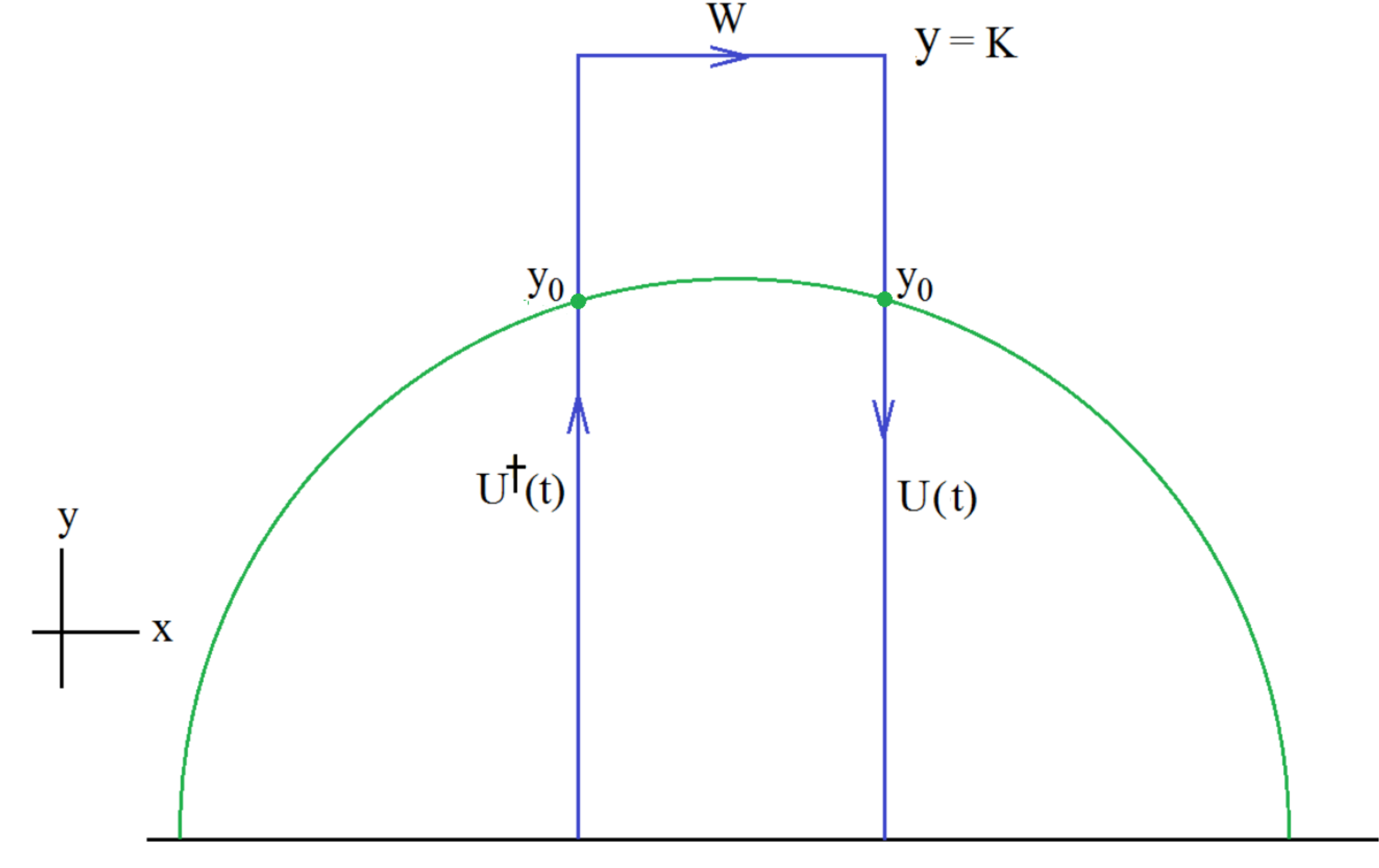}
\caption{ The $\c_2$-geometry may be drawn as a rectangular piece of the Poincare plane. The green semicircle is a geodesic. }
\label{7}
\end{center}
\end{figure}

With the assumption of uniform negative curvature, the $\c_2$-geometry may be drawn as a piece of the Poincare plane. This is shown in figure \ref{7}.

The metric has the form,

\be
dl^2 = \frac{K^2}{4y^2}(dx^2 + dy^2).
\label{2metric}
\ee

\subsection{Complexity Calculation}
Let us start by calculating the complexity of $U(t).$ According to our assumptions the complexity should be given by $Kt/2l,$ and should be equal to the geodesic length of the vertical segment representing $U(t).$

The vertical geodesic distance between the top of the rectangle (location of $W$) and the point $y_0$ is given by

\be
 \frac{K}{2} \int_{y_0}^K \frac{dy}{y} =  \frac{K}{2} \log{\frac{K}{y_0}}.
\ee

This allows us to identify the relation between $t$ and $y_0,$

\be
t=l \log{\frac{K}{y_0}}.
\ee

The naive expectation for the complexity of $W(t)$ is just the sum of the geodesic lengths of the three blue segments in figure \ref{7}.
Adding the two vertical lengths and ignoring the small contribution of $W$, the naive complexity would be,

\be
\c_W = K  \log{\frac{K}{y_0}} =\frac{Kt}{l}  \ \ \ \ \ \ \ (\rm  naive \it).
\label{naive2}
\ee

But \ref{naive2} is based on a wrong identification of the complexity with distance along the broken curve. According to the arguments of \cite{Nielsen}\cite{Dowling} the complexity should be identified as the
 geodesic distance between the two points at $y_0.$ Thus we must find the horizontal geodesic connecting corresponding points on the vertical geodesics.
  This is easy to do since all geodesics on the Poincare plane are circles centered on the $x$ axis. The geodesic passing through the two points at $y_0$ is shown in green. It is given by,

\bea
x^2 + y^2  \eq L^2 \cr \cr
L^2 \eq 1 + y_0^2
\eea

Its length is easily calculated  from the metric \ref{2metric},

\bea
\c_W \eq \frac{K}{2} \log{\frac{L+1}{L-1}} \cr \cr
\eq \frac{K}{2} \log{\frac{{\sqrt{e^{2t/l} +K^2}}+e^{t/l}}{{\sqrt{e^{2t/l} +K^2}}-e^{t/l}  }}
\label{complexity from geometry}
\eea

From \ref{complexity from geometry} it is easily seen that $\c_W$ behaves like

\bea
\c_W(t) &\to&  e^{t/l}   \ \ \ \ \ \ (t<l\log{K}) \cr \cr
\c_W(t) &\to& \frac{K}{l} (t - l\log{K})   \ \ \ \ \ \  \ \   (t>l\log{K})
\label{result for c}
\eea
This is very similar to the results from the quantum circuit analysis summarized in figure \ref{5}. It is significant evidence that the complexity properties of precursors are well described by a $\c_2$-geometry with the simple property of uniform negative curvature.

The shift from $t$ to $(t-l\log K)$ in the second equation of \ref{result for c} is the switchback effect of \cite{Stanford:2014jda} whose importance we discussed earlier. In $\c$-geometric context it occurs because complexity is defined by the minimum geodesic distance and not the length of the original path consisting of $U(t),$ $W,$ and $U(-t).$

\subsection{Size of Precursor}

Now let us consider the size function $s_w.$ We don't have a first-principles way to connect the size with $\c$-geometry, but inspired by equation \ref{s=dc/dt} we may write

\bea
2s_w(t) \eq \frac{d}{dt} K \log{\frac{L+1}{L-1}} \cr \cr
\eq -\frac{2K}{L^2-1}\frac{dL}{dt}
\eea

Using $$L^2= y_0^2 +1$$ and $t=l (\log{K}-\log{y_0})$
we find

\be
s_w(t) = \frac{K}{L}
\label{s=2K/L}
\ee

This can also be written as

\be
s_w(t) = \frac{K}{\sqrt{1+ K^2 e^{-2t/l}}}
\label{s from line}
\ee

\begin{figure}[h!]
\begin{center}
\includegraphics[scale=.3]{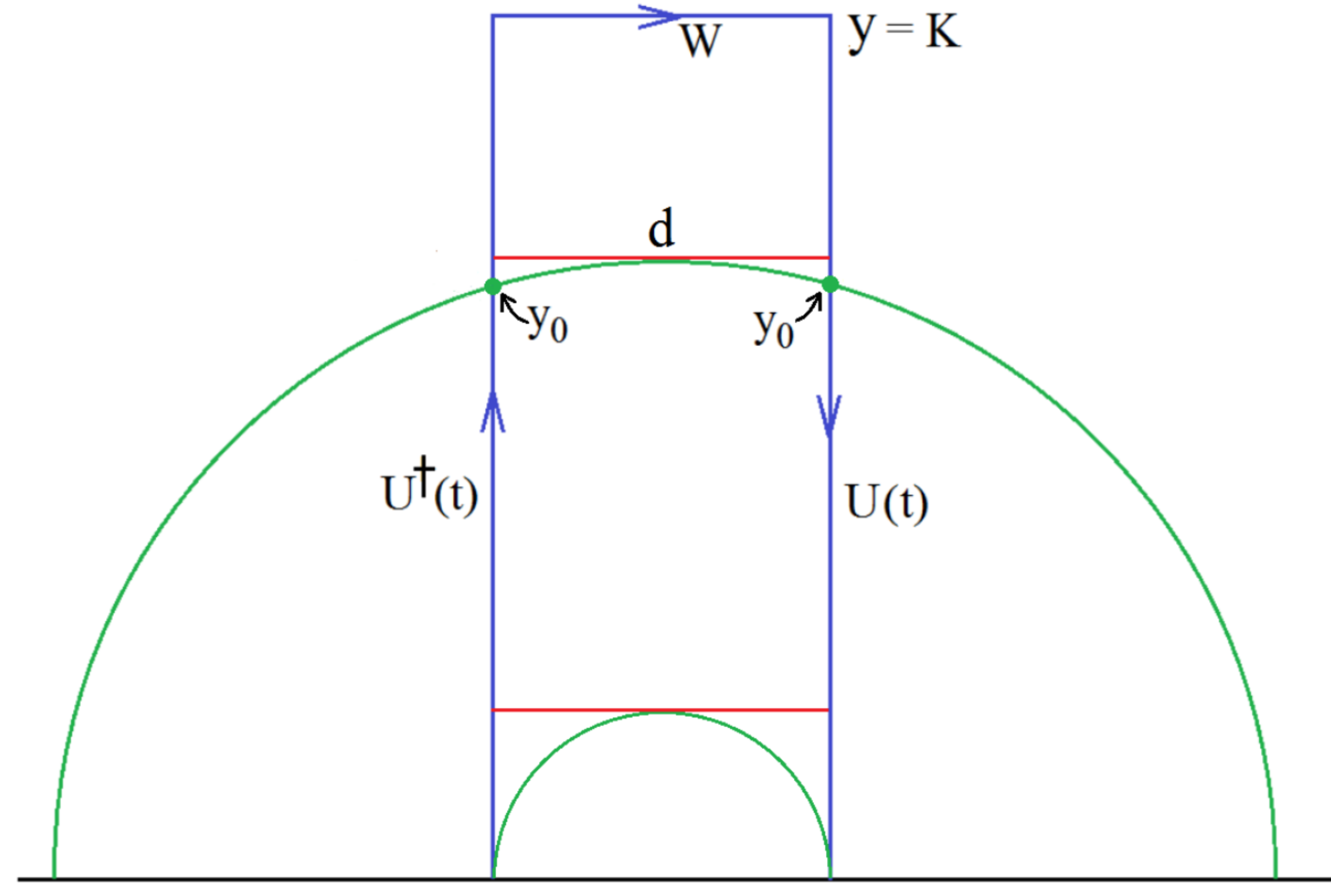}
\caption{The red lines are not geodesics. Their metric length represents the size of the precursor at time $t=\log{K/y_0}.$ }
\label{8}
\end{center}
\end{figure}

Equation \ref{s=2K/L} is a remarkably simple result which has the geometric interpretation of the
distance along a line of constant $y$ tangent to the top of the circle $x^2+y^2 = L^2$ as shown in figure \ref{8}. We'll denote it by the symbol $d(t).$ It should be understood that the time $t$ does not refer to the points at which the red segment intersects the vertical geodesics; it refers to the points in the figured labeled $y_0.$

The red segments in  figure \ref{8}  are not geodesics, and its not clear what is special about them from the geometric viewpoint. We would like to suggest an alternate interpretation. We can get a hint from dimensional analysis, where the dimensions refer to $\c$-geometry, not space-time. From the form of the metric in \ref{2metric} we see that the constant $K$ has units of length. Both complexity and size scale with one power of $K$ and are therefore also lengths in ``complexity units".

On the other hand the double sum $\s(t)$ in \ref{double sum} scales like $K^2$ and is therefore an area in complexity units. It is of course just the product $sK.$

\be
\s(t) = d(t)K.
\label{s=dK}
\ee

It follows from the properties of the Poincare plane that the  area of the column above the red segment, from the red segment to $y=\infty,$ is also $d(t)\frac{K}{2}.$ Let us call that area $\a'.$
Equation \ref{s=dK} becomes

 \be
 \s = 2\a'.
 \ee

\begin{figure}[h!]
\begin{center}
\includegraphics[scale=.3]{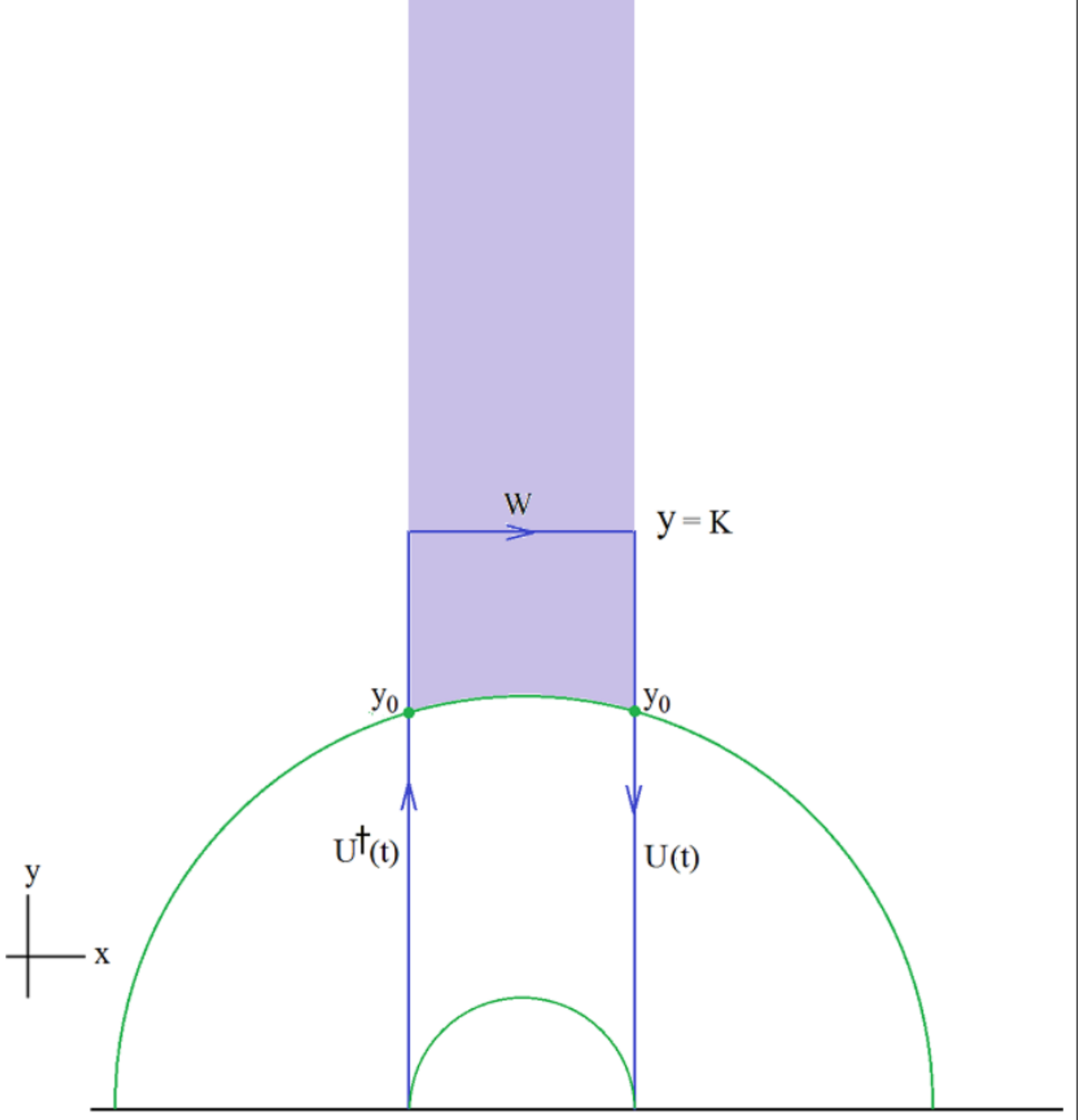}
\caption{The area of the shaded region is approximately equal to the function $\s(t).$ }
\label{9}
\end{center}
\end{figure}

 Still, the lack of a clear geometric meaning to the red line, and the area above it, remain unsatisfying. A more attractive quantity is the area above the green geodesic shown in figure \ref{9}. Call it $\a.$   How much different are $\a'$ and $\a?$ Not very. They differ by the area of the two small triangular regions between the two curves. One might worry that these areas grow proportionally larger than $\a'$ when $y_0$ gets small, but this doesn't happen. The maximum ratio of $\a$ to $\a'$ is,

 \be
\lim_{t\to\infty} \frac{a}{a'} = \frac{\pi}{2}
 \ee
For times  less than the scrambling time the ratio is close to unity.

We find it  tempting to speculate that in the realistic Hamiltonian evolution that an appropriately defined version of $\s$ is accurately represented by the area $\a.$

The calculation of $\a$ is straightforward. It is given by,

\bea
\a \eq\frac{ K^2}{4} \int_{-1}^{1} dx \int_{\sqrt{L^2-x^2}}^{\infty} \frac{dy}{y^2} \cr \cr
\eq
 \frac{K^2}{2} \left[
\arcsin\frac{1}{\sqrt{1+y_0^2}}
\right]
\eea

or in terms of an equivalent $s(t),$

\be
s_{w}(t) = K \left[
\arcsin\frac{1}{\sqrt{1+  K^2 e^{-2t/l}      }}
\right]
\label{s from  area}
\ee

For small time,  one finds the area having the form,

\be
\a = \frac{K}{2} e^{t/l}
\label{area exp}
\ee

For large $t$ the area has the limiting value,

\be
\a \to \frac{\pi}{4} K^2
\label{area const}
\ee

compared with $a'$ which has the limiting  behavior

\be
\a' \to \frac{ K^2}{2}.
\label{area to const}
\ee

We have constructed three functions to represent the size $s_w(t),$ namely
\ref{s from epidemic},  \ref{s from line}, and \ref{s from  area}. The first came from analyzing the behavior of random quantum circuits, and the other two from $\c_2$-geometry. It is interesting to compare them. We can do this best by first changing the $t$ variable to $u,$

\be
u = \frac{1}{l}(t-t_{\ast}) = \frac{1}{l}(t-t_{\ast}) = \frac{t}{l} -\log{K}.
\ee

Each of the three functions has the form
\be
s = K F(u).
\ee

For the three cases \ref{s from epidemic},  \ref{s from line},  \ref{s from  area} the function  $F$ takes the form,

\be
F_1 = \frac{e^u}{1+e^u}
\label{f1}
\ee
for  \ref{s from epidemic};

\be
F_2 =\frac{e^u}{\sqrt{1+e^{2u}}}
\label{f2}
\ee
for \ref{s from line};  and

\be
F_3= \frac{2}{\pi}\arcsin{ \frac{e^u}{\sqrt{1+e^{2u}}}  }
\label{f3}
\ee
for \ref{s from  area}. The factor of $2/\pi$ was included in $F_3$ in order to normalize the asymptotic value to $1.$

\begin{figure}[h!]
\begin{center}
\includegraphics[scale=.3]{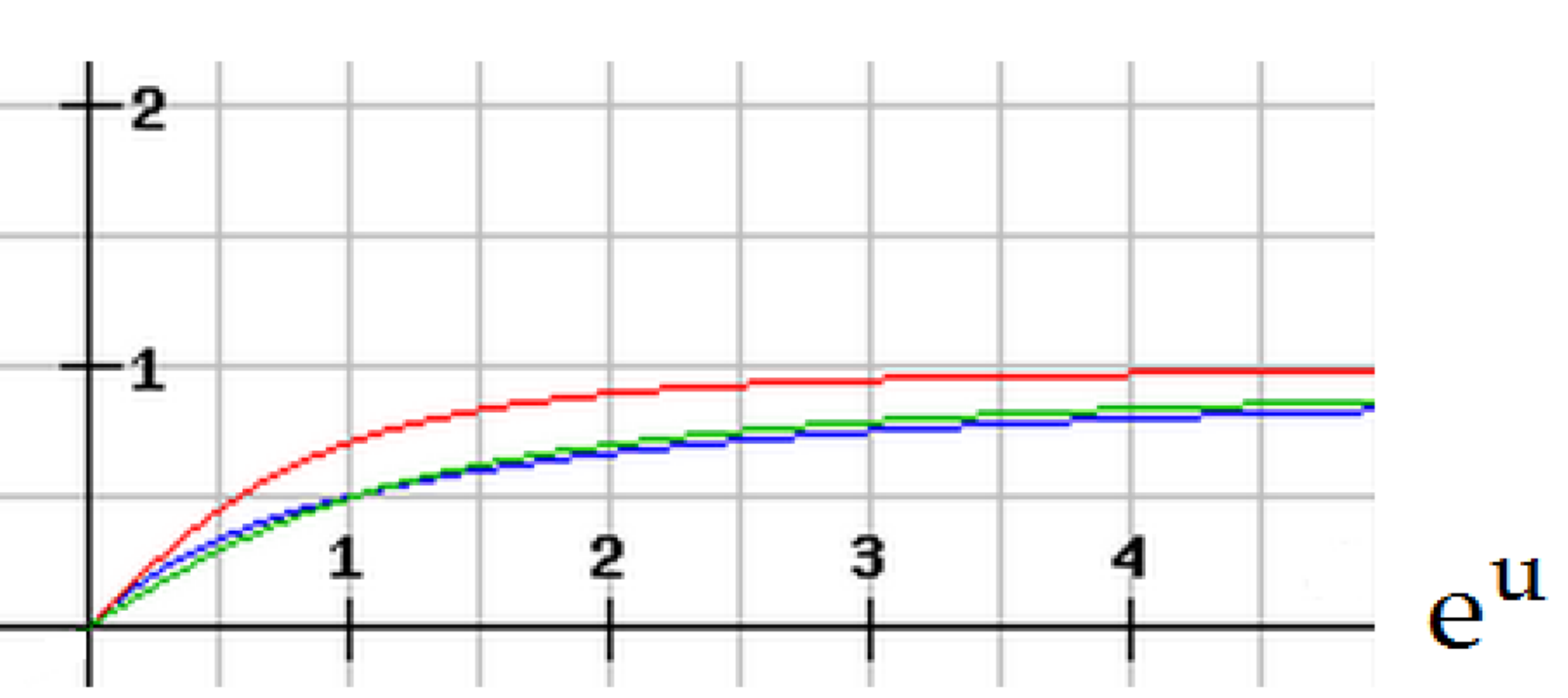}
\caption{ The three functions \ref{f1}, \ref{f2}, and \ref{f3} are plotted in blue, green, and red respectively. The horizontal axis is $e^u.$}
\label{10}
\end{center}
\end{figure}

The three functions are plotted in figure \ref{10}

\begin{figure}[h!]
\begin{center}
\includegraphics[scale=.3]{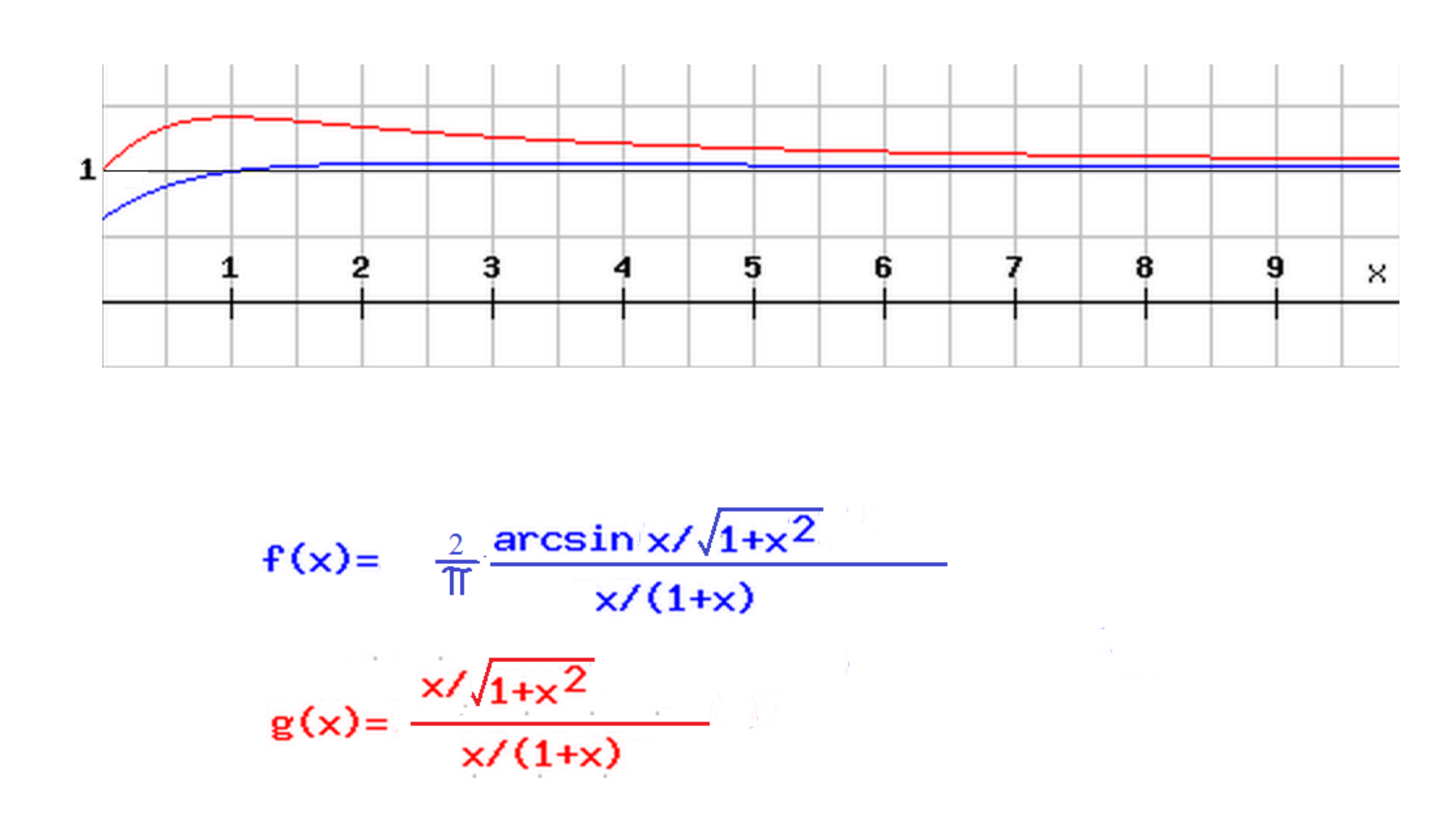}
\caption{ Ratios plotted}
\label{11}
\end{center}
\end{figure}

While the functions are generally similar, \ref{f3} and \ref{f1} are extremely close over most of the range, while \ref{f2} differs substantially. We can see the same thing by plotting the ratios in in figure \ref{11} ( $F_2/F_1$ (red) and $F_3/F_1$ blue).

It is not clear to us why $F_3$ is a better fit to $F_1$ than $F_2.$

\section{Questions}

There are many questions that this paper leaves unanswered. We will list a few.

\bi
\item When Hawking radiation is accounted for, the bridge-to-nowhere will become a bridge-to-somewhere, the somewhere being the Hawking radiation that purifies the state of the black hole \cite{Maldacena:2013xja}. What are the properties of such a bridge? Can it be described by a quantum circuit and its close relative, a tensor network \cite{Hartman:2013qma}\cite{Stanford:2014jda}?

    \item The trajectory $U(t)$ swept out by an easy Hamiltonian is a geodesic, but not necessarily the shortest geodesic. For some period of time it will be the shortest, but that cannot last forever. Since complexity is bounded by by exponential in $K,$ shorter geodesics must come into play by the classical recurrence time $\sim e^K.$ This happens when cut points occur. The occurrence of a cut point may signal the breakdown of the classical general relativistic description of the ERB.

        How long does it take for the first cut point to occur? Our guess is that it takes the classical recurrence time but there is no strong evidence for this.

        \item We conjectured a very simple $\c_2$-geometry to describe precursor complexity, and it appears to capture the expected properties of precursors rather well. But there was no first-principles derivation of it from the $\c$-geometry of \cite{Nielsen}\cite{Dowling}.
        Can we derive its main properties directly from definition of $\c$-geometry and the properties of black holes such as fast scrambling? In particular can we demonstrate that the curvature of $\c_2$ is of order $-1/K^2?$

          \item We restricted our discussion to black holes of radius equal to the ADS radius $l.$ How much of the analysis generalizes to larger black holes? This question will partly be answered in a forthcoming paper \cite{RSS}.

            \item Perhaps the biggest issue is how to understand the relation between general relativity and complexity. GR gives a description of ERBs, and  certain aspects of the geometry closely parallel the growth of quantum complexity including the details of precursors and switchbacks. How much more of GR can be found hidden in the study of quantum complexity? We expect much more.

\ei

\section*{Acknowledgements}

We thank Patrick Hayden and   Nathaniel Thomas for explaining the work of Dowling and Nielsen. We also thank Michael Nielsen for correcting some misconceptions.

Special thanks goes to Dan Roberts and Douglas Stanford for sharing their insights about the behavior of $s$ and $\c$ from the study of quantum circuit models.

Support for this research came through NSF grant Phy-1316699 and the Stanford Institute for Theoretical Physics.

\appendix
\section{Geometry of ERB1}

In this appendix we will calculate the functional form of the ERB1 for the one-sided version of a BTZ black hole.

\subsection{Metric}
Consider a one-sided AdS black hole formed by gravitational collapse of a spherical shell of null matter. The metric is
\be
ds^2=-f(r)dt^2+\frac{dr^2}{f(r)}+r^2d\phi^2,
\ee
where
\be
f(r)=\frac{r^2}{l^2}+1
\ee
before the shock wave (empty AdS), and
\be
f(r)=\frac{r^2}{l^2}-1
\ee
after the shock wave (BTZ).

\begin{figure}[h!]
\begin{center}
\includegraphics[scale=.3]{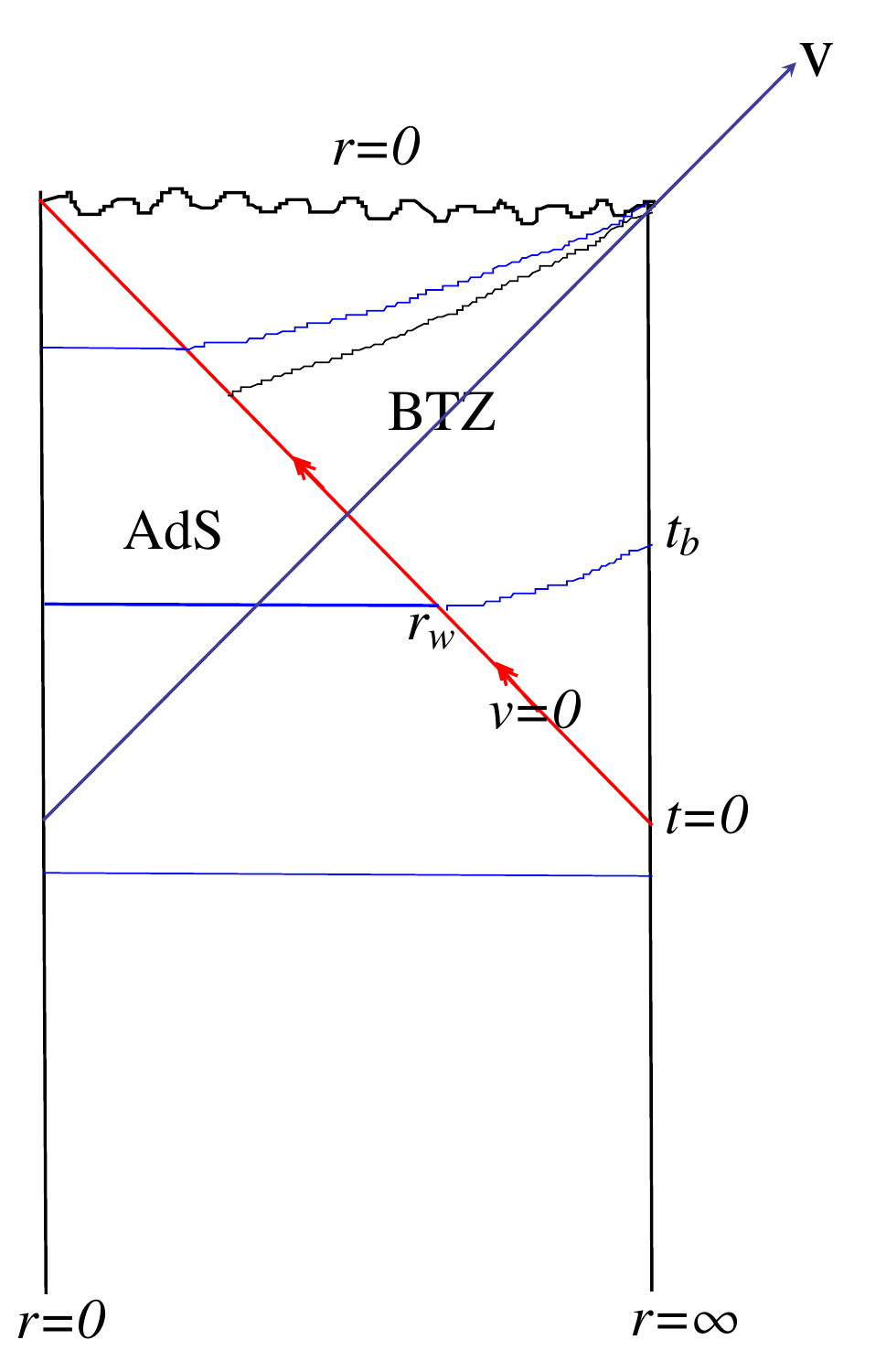}
\caption{The ERB1 with points and regions labeled. }
\label{12}
\end{center}
\end{figure}

The $r$-coordinates, which characterizes the proper area around the center, is continuous across the shock wave. We also choose $t$-coordinates to be continuous along the boundary $r=\infty$ as a convention.

This coordinate doesn't cover the horizon in BTZ part. Passing to null coordinates, let
\be
dv=dt+\frac{dr}{f(r)},
\ee
then
\be
ds^2=-f(r)dv^2+2drdv+r^2d\phi^2.
\ee

More specific, in AdS part, let
\be
v=t+l(\arctan\frac{r}{l}-\frac{\pi}{2}),
\ee
and in BTZ part, let
\be
v=t+\frac{l}{2}\log\left|\frac{r-l}{r+l}\right|,
\ee
then the matching condition along the shock wave requires both $r$ and $v$ coordinates to be continuous. If the shock wave comes in at boundary time $t=0$, then it lies at $v_w=0$.

\subsection{Extremal Surface}
We look for extremal spacelike surface ending at boundary time $t_b$. When $t_b<0$, the surface is simply a disk in pure AdS with $t=$ constant. Its area is
\be
\frac{A(t<0)}{2\pi l^2}=\int_0^{r_b}dr\frac{r}{\sqrt{\frac{r^2}{l^2}+1}}=\frac{r_b}{l}
\ee
where $r_b$ is the boundary radius cutoff.

 After the shockwave, the surface will contain two parts, one disk inside pure AdS, and another part inside BTZ. We parametrize the surface by a function $r=r(v)$. Then the area element is
 \be
 \frac{dA}{2\pi}=r\sqrt{2r'-f}dv,
 \ee
where $r'$ is the derivative of $r$ with respect to $v$.
We can consider $L=r\sqrt{2r'-f}$ as a lagrangian, $v$ as time, and $r$ as canonical coordinate. Then the conjugate momentum is
\be
p_r=\frac{\partial L}{\partial r'}=\frac{r}{\sqrt{2r'-f}}.
\ee
In both the AdS and the BTZ region, the Lagrangian is independent of $v$ (time), so the corresponding energy is separately conserved
\be
E=r'p_r-L=\frac{f-r'}{\sqrt{2r'-f}}.
\ee

Look at the pure AdS part first. Smoothness at $r=0$ requires $0=\frac{\displaystyle dt}{\displaystyle dr}=\frac{\displaystyle dv}{\displaystyle dr}-f$. So $r'=f$ at $r=0$, and therefore the energy $E_{AdS}=0$. Then it follows that $r'=f$ on the entire surface inside AdS, which just says it's a disk $t=$ constant. Let's label the surface by the $r$-coordinates of its intersection with the shockwave $r_w$, then
\be
\label{surfaceAdS}
\frac{t}{l}=\frac{\pi}{2}-\arctan\frac{r_w}{l}.
\ee
Its area is
\be
\frac{A_{AdS}}{2\pi l^2}=\sqrt{\frac{r_w^2}{l^2}+1}.
\ee
At the intersection point, the momentum is
\be
p_r(v=0,r=r_w)=\frac{\displaystyle r_w}{\displaystyle \sqrt{\frac{r_w^2}{l^2}+1}}.
\ee

What's the matching condition across the shock wave? In the corresponding mechanics problem, there is a change of Lagrangian at time $v=0$. First we must require the coordinate $r$ to be continuous. Also, if we vary the end point along that $v=0$ time slice, when the equation of motion is satisfied, the change of the action is $p_r\delta r$. So the local extremal condition requires that momentum $p_r$ is also continuous across the shockwave.

We get that just across the shock wave in the BTZ part,
\bea
&2r'-(\frac{r_w^2}{l^2}-1)=\frac{r_w^2}{l^2}+1,\\
&r'=\frac{r_w^2}{l^2}
\eea
The energy in the BTZ part is
\be
E_{BTZ}=-\frac{r_w}{\sqrt{\frac{r_w^2}{l^2}+1}}.
\ee

To simplify notation, let's parametrize the crossing point by $\tan\theta=\frac{\displaystyle r_w}{\displaystyle l}$, then
\be
E_{BTZ}=-l\sin\theta=\frac{r(f-r')}{\sqrt{2r'-f}}.
\ee

Solving this, and picking up the appropriate root, we get
\be
\label{surfaceBTZ}
v(r)=\int_{l\tan\theta}^r\frac{dr'}{\displaystyle \sqrt{\frac{r'^2}{l^2}-1+\frac{l^2}{r'^2}\sin^2\theta}\left(\sqrt{\frac{r'^2}{l^2}-1+\frac{l^2}{r'^2}\sin^2\theta}+\frac{l}{r'}\sin\theta\right)}.
\ee
This gives the shape of the extremal surface in BTZ part parametrized by $\theta$.

Its area is
\be
\label{area}
\frac{A_{BTZ}}{2\pi l^2}=\int_{l\tan\theta}^{r_b}dr\frac{\displaystyle r}{\displaystyle l^2\sqrt{\frac{r^2}{l^2}-1+\frac{l^2}{r^2}\sin^2\theta}}.
\ee

We still need to relate the parameter $\theta$ to the boundary time $t_b$. At boundary, $v_b=t_b$, so
\be
\label{time}
t_b=\int_{l\tan\theta}^{\infty}\frac{\displaystyle dr}{\displaystyle \sqrt{\frac{r^2}{l^2}-1+\frac{l^2}{r^2}\sin^2\theta}\left(\sqrt{\frac{r^2}{l^2}-1+\frac{l^2}{r^2}\sin^2\theta}+\frac{l}{r}\sin\theta\right)}.
\ee
\\
To consider small time behavior, we can expand everything to linear order in $\frac{\displaystyle \pi}{\displaystyle 2}-\theta$, then we find
\be
\frac{A}{2\pi l^2}=\frac{r_b}{l}+\frac{t_b}{l}+\mathcal{O}\left(\frac{t_b}{l}\right)^2.
\ee
\\
At large time, look at the equation (\ref{time}).
$\theta$ decreases as we increase $t_b$. The expression inside the square root stays positive and the integral is finite until we reach $\sin\theta=\frac{1}{2}$. At $\sin\theta=\frac{1}{2}$, i.e., $r_w=\sqrt{\frac{1}{3}}l$, the integral is infinite as $r$ goes from $\sqrt{\frac{1}{3}}l$ to $\sqrt{\frac{1}{2}}l$, which means that the surface ends at intersection of the boundary and singularity. This is the limiting surface at large time. In this case the expression inside the square root is a complete square, and we can do the integral of (\ref{surfaceBTZ}) analytically.
\be
\frac{v}{l}=\log\left(\frac{1-\frac{r}{l}}{1+\frac{r}{l}}\right)+\frac{1}{\sqrt 2}\log\left(\frac{\frac{1}{\sqrt 2}+\frac{r}{l}}{\frac{1}{\sqrt 2}-\frac{r}{l}}\right)-\log\left(\frac{1-\frac{1}{\sqrt 3}}{1+\frac{1}{\sqrt 3}}\right)-\frac{1}{\sqrt 2}\log\left(\frac{\frac{1}{\sqrt 2}+\frac{1}{\sqrt 3}}{\frac{1}{\sqrt 2}-\frac{1}{\sqrt 3}}\right).
\ee

This limiting surface lies completely inside the horizon. In $t$, $r$ coordinates, it's
\be
\label{limiting surface}
\frac{t}{l}=\frac{1}{2}\log\left(\frac{1-\frac{r}{l}}{1+\frac{r}{l}}\right)+\frac{1}{\sqrt 2}\log\left(\frac{\frac{1}{\sqrt 2}+\frac{r}{l}}{\frac{1}{\sqrt 2}-\frac{r}{l}}\right)+\text{constant}.
\ee

In the black hole region $r\sqrt{-f(r)}$ is maximized on the surface $r =r_m = \sqrt{\frac{1}{2}}l.$ The limiting surface we found in (\ref{limiting surface}) lies entirely inside $r=r_m$, and approaches it exponentially in $t$ coordinate as we go to the boundary.
\be
\frac{r}{l}\longrightarrow\frac{1}{\sqrt 2}-\text{const}\cdot \exp\left(\displaystyle-\frac{\sqrt 2 t}{l}\right).
\ee

At large $t_b$, the growth of the surface is controlled by the $r=r_m$ surface.
\be
\frac{A}{2\pi l^2}\sim \frac{r_m}{l}\sqrt{1-\frac{r_m^2}{l^2}}\frac{t_b}{l}=\frac{t_b}{2l}.
\ee

In more details, consider time and area integrals (\ref{time}) and (\ref{area}). $r$ goes from $r_w$ to the boundary infinity. At large $t_b$, $\sin\theta$ becomes very close to $\frac{1}{2}$, and the denominators of both integrals become very small near $r=r_m=\sqrt{\frac{1}{2}}l$. We can divide the integrals into three regions, (I) $r$ goes from $r_w=l\tan\theta$ to $r_m-\Delta$, (II) from $r_m-\Delta$ to $r_m+\Delta$, and (III) from $r_m+\Delta$ to $r_b\longrightarrow\infty$, where $\frac{\Delta}{r}$ is small. At large boundary time $t_b$, we expect the surface to get very close to the limiting surface. How close? The crossing point on the shockwave approaches its limit exponentially in $t_b$,
\be
\sin\theta=\frac{1}{2}+\text{const}\cdot \exp\left(-\frac{\sqrt 2 t_b}{l}\right).
\ee
Also, in the time integral (\ref{time}), contributions from region (I) and (III) stay almost constant ($\Delta$-dependent), while region (II) gives most contribution, increases like $t_b$, i.e, the surface spends most of its time around the $r=r_m$ surface, roughly half inside and half outside. The behavior of the area integral (\ref{area}) is similar. The contributions from region (I) and (III) are almost constant, while the contribution of region (II) goes like $\frac{\displaystyle t_b}{\displaystyle 2l}$. This is where the large time linear behavior comes out. \\

Here is a picture of the large time surface. As $t_b$ increases, region (I) stays almost exactly at the limiting surface and gives a constant area. Region (II) stays around the Hartman-Maldacena surface and gets longer, giving an area linear in $t_b$. The position of region (III) changes since it's pushed out by region (II). It roughly looks like
\be
\label{surfacetail}
v= t_b+\frac{\displaystyle l}{\displaystyle \sqrt 2}\log\left(\frac{\displaystyle \frac{r}{l}-\frac{1}{\sqrt 2}}{\displaystyle\frac{r}{l}+\frac{1}{\sqrt 2}}\right).
\ee
But its contribution to area remains again almost constant. We can see this by evaluating the time integral (\ref{time}) in region (III), or, we can see this directly from the above expression (\ref{surfacetail}). It's essentially the same surface got pushed out in $v$-direction as $t_b$ increases. Its contribution to area will be divergent, goes like $\frac{\displaystyle r_b}{\displaystyle l}$, but is $t_b$-independent anyway. \\

\end{document}